# galpy: A python LIBRARY FOR GALACTIC DYNAMICS

Jo Bovy[1,2]

Institute for Advanced Study, Einstein Drive, Princeton, NJ 08540, USA; bovy@ias.edu

## ABSTRACT

I describe the design, implementation, and usage of galpy, a python package for galactic-dynamics calculations. At its core, galpy consists of a general framework for representing galactic potentials both in python and in C (for accelerated computations); galpy functions, objects, and methods can generally take arbitrary combinations of these as arguments. Numerical orbit integration is supported with a variety of Runge-Kutta-type and symplectic integrators. For planar orbits, integration of the phase-space volume is also possible. galpy supports the calculation of action-angle coordinates and orbital frequencies for a given phase-space point for general spherical potentials, using state-of-the-art numerical approximations for axisymmetric potentials, and making use of a recent general approximation for any static potential. A number of different distribution functions (DFs) are also included in the current release; currently these consist of two-dimensional axisymmetric and non-axisymmetric disk DFs, a three-dimensional disk DF, and a DF framework for tidal streams. I provide several examples to illustrate the use of the code. I present a simple model for the Milky Way's gravitational potential consistent with the latest observations. I also numerically calculate the Oort functions for different tracer populations of stars and compare it to a new analytical approximation. Additionally, I characterize the response of a kinematically-warm disk to an elliptical $m = 2$ perturbation in detail. Overall, galpy consists of about 54,000 lines, including 23,000 lines of code in the module, 11,000 lines of test code, and about 20,000 lines of documentation. The test suite covers 99.6 % of the code. galpy is available at http://github.com/jobovy/galpy with extensive documentation available at http://galpy.readthedocs.org/en/latest .

*Subject headings:* galaxies: general — galaxies: kinematics and dynamics — Galaxy: fundamental parameters — Galaxy: structure — methods: numerical — stellar dynamics

## 1. INTRODUCTION

Galactic dynamics is an old and venerable subject in astrophysics and a mainstay in the Astronomy and Astrophysics graduate-school curriculum. While many good textbooks are devoted to the topic (e.g., Binney & Tremaine 2008), very few software tools designed to aid in galactic-dynamics computations are currently available (with the notable exception of *N*-body codes, for which there are many publicly accessible packages, e.g., Springel 2005). One of the only such software packages is the NEMO Stellar Dynamics Toolbox (Teuben 1995), which is a collection of various programs to set up, evolve, and analyze *N*-body systems, but that can also be used for more basic computations involving orbits in galactic potentials. NEMO follows a UNIX-like workflow of pipes and filters and various programs can be strung together in the terminal to create very powerful workflows.

In this paper, I introduce a new, modern software toolbox for galactic dynamics that is wholly focused on the dynamics and distributions of orbits in external gravitational potentials: galpy. The galpy package is largely written in the python programming language and combines the flexibility and ease-of-use of a high-level, object-oriented language with the speed of the lower level C language to resolve speed bottlenecks. The basic functionality of galpy only depends on the scientific software python libraries numpy (Oliphant 2006), scipy (Jones et al. 2001), and matplotlib (Hunter 2007) and is therefore straightforward to install. galpy's interface is designed to be simple, intuitive, and easily extensible.

At its core galpy contains Potential objects and Orbit objects that can represent a variety of galactic potentials and orbits in those potentials. galpy contains a large number of functions to characterize arbitrary combinations of basic potentials contained in galpy as well as user-specified potentials, to integrate and characterize orbits in these potentials, and to work with distributions of orbits. While being generally useful in the study of galactic dynamics, galpy's powerful framework will be especially useful in aiding in the interpretation of the exquisite kinematic data from the astrometric Gaia satellite scheduled to appear soon (Perryman et al. 2001).

This paper describes the basic functionality of galpy contained in the v1.0 release. As with most software, galpy's development is ongoing and features are being added regularly. galpy is an open-source code that is being developed under the git version-control system on GitHub:

http://github.com/jobovy/galpy

and the latest documentation can be found at

http://galpy.readthedocs.org/en/latest/ .

The outline of this paper is as follows. In § 2, I give a general overview of the structure and design of galpy. In § 2.2 I discuss the units that galpy uses and how to convert to and from these units. § 3 introduces the Potential object, its various methods, and how it can be used and extended. In § 3.4, I describe a few non-axisymmetric potentials that are implemented and in

[1] Hubble Fellow
[2] John Bahcall Fellow

§ 3.5 I introduce a fiducial model for the gravitational potential of the Milky Way that is included in `galpy` and that can be used when interpreting Milky-Way data. In § 4, I present `Orbit` objects and their instance methods that can be used to extract orbital characteristics. Orbital actions, frequencies, and angles are particularly informative attributes of an orbit and their calculation within the `galpy` framework is discussed in detail in § 5: § 5.2 describes the action–angle calculations for the isochrone potential and other spherical potentials, § 5.3 conveys how two approximations for axisymmetric potentials are implemented, and § 5.5 discusses the implementation of a general method for calculating actions, angles, and frequencies.

I describe various distribution functions (DFs) contained in `galpy` in § 6: this includes two-dimensional DFs for the mid-plane of a disk galaxy as well as a three-dimensional DFs that are based on actions. The two-dimensional DFs include a module for non-axisymmetric DFs and I give examples of the response of a kinematically-warm stellar disk to elliptical perturbations as well as to the influence from a central bar in § 7. Some aspects related to the structure of the `galpy` codebase and its development are described in § 8.1 and the test suite and corresponding test coverage statistics are presented in § 8.2. Finally, in § 9, I look ahead to upcoming additions to the `galpy` codebase. An Appendix discusses various coordinate transformations that are implemented in `galpy`.

Self-contained code to reproduce all of the plots in this paper except for those in § 7 (which require a very large amount of time to run) is available in a `git` repository associated with this particular paper. This repository is separate from the `galpy` repository and can be found on `GitHub` as well at

http://github.com/jobovy/galpy-paper-figures .

## 2. GENERAL `galpy` OVERVIEW

### 2.1. *Package structure*

The overall structure of the `galpy` package is illustrated in Figure 1, where the package and its main subpackages are imported in a typical `python` session. These subpackages have fairly obvious names: `galpy.potential` contains classes and functions related to gravitational potentials, `galpy.orbit` consists of the `Orbit` class and all of the functionality related to it, `galpy.actionAngle` has specialized classes and functions for dealing with action–angle coordinates, and `galpy.df` is composed of classes and functions related to distribution functions. The `galpy.util` subpackage collects a large variety of utility functions related to plotting, coordinate transformations, and unit conversions.

As described in more detail in § 3, specific potentials are included as sub-classes of a general `galpy.potential.Potential` class under `galpy.potential`. For example, `galpy.potential` contains a Navarro-Frenk-White potential (NFW; Navarro et al. 1997), which can be imported and used as

```
from galpy.potential import NFWPotential
np= NFWPotential(normalize=1.)
```

This initialization is explained further below. `Potential`

```
1  import galpy
2  import galpy.potential
3  import galpy.orbit
4  import galpy.actionAngle
5  import galpy.df
6  import galpy.util
```

Fig. 1.— Importing `galpy` and its subpackages to illustrate the overall structure of `galpy`.

objects form the basis of almost all of the `galpy` functionality.

The `Orbit` object is another central part of `galpy`'s infrastructure. It can be set up in a variety of flexible ways that are described in § 4. For the general overview, we note that `Orbit` instances can be initialized as

```
from galpy.orbit import Orbit
o= Orbit(vxvv=[1.,0.1,1.1,0.1,0.02,0.])
```

which sets up an `Orbit` instance `o` that represents the initial conditions of an orbit. This orbit can then be integrated and its orbital characteristics can be evaluated using member functions.

The `galpy.actionAngle` module contains classes that represent different ways of calculating action–angle coordinates. These are always set up for a specific potential and can then be used to calculate actions, frequencies, and angles for specific orbits. A quick example that illustrates this using the NFW potential and `Orbit` instance defined above is

```
from galpy.actionAngle import actionAngleSpherical
aA= actionAngleSpherical(pot=np)
aA(o)
(array([ 0.00980542]), array([ 1.1]),
array([ 0.00553155]))
```

where the output are the radial, azimuthal, and vertical action of the orbit. The `galpy.actionAngle` module is described further in § 5.

The `galpy.df` subpackage contains a few classes that represent different DFs. These typically depend on `Potential` and `Orbit` instances and in some cases use `actionAngle` instances when the DF depends on the actions. For example, one action-based DF that is included in `galpy` is the quasi-isothermal DF (Binney 2010; Binney & McMillan 2011), which can be used as follows

```
from galpy.df import quasiisothermaldf
qdf= quasiisothermaldf(1./3.,0.2,0.1,1.,1.,
                       pot=np,aA=aA)
qdf(o)
array([ 61.57476085])
```

where the output is the value of the DF for this orbit. The various DF classes currently included in `galpy` are discussed in § 6.

Finally, the utility subpackage `galpy.util` contains plotting utilities in `galpy.util.bovy_plot`, coordinate-transformation routines in `galpy.util.bovy_coords`, and unit-conversion functions in `galpy.util.bovy_conversion`. The latter are presented in the next subsection, as they relate to the system of units used preferentially by `galpy`. The coordinate transformations are those between equatorial and Galactic coordinates, Galactic coordinates and

```
1 import galpy.util.bovy_conversion as conversion
2 print conversion.force_in_pcMyr2(220.,8.) #pc/Myr^2
3 6.32793804994
4 print conversion.dens_in_msolpc3(220.,8.) #Msolar/pc^3
5 0.175790330079
6 print conversion.surfdens_in_msolpc2(220.,8.) #Msolar/pc^2
7 1406.32264063
8 print conversion.mass_in_1010msol(220.,8.) #10^10 Msolar
9 9.00046490005
10 print conversion.freq_in_Gyr(220.,8.) #1/Gyr
11 28.1245845523
12 print conversion.time_in_Gyr(220.,8.) #Gyr
13 0.0355560807712
```

Fig. 2.— Units in galpy: Illustration of the use of the galpy.util.bovy_conversion subpackage for conversion between physical units where the circular velocity is 220 km s$^{-1}$ at $R = 8$ kpc and galpy's *natural units* where velocities and positions are scaled by these factors. Physical units are obtained by multiplying output in natural units by these factors.

Galactocentric coordinates, and uncertainty propagation between some of these coordinate systems. These are discussed in some detail in the Appendix. The plotting routines are simple wrappers of matplotlib plotting functions; these are documented at

http://galpy.readthedocs.org/en/latest/reference/bovyplot.html .

## 2.2. *Units in* galpy

While generally galpy does not care about the units of its inputs (as long as they are consistent), some of the functionality will work best when using *natural units*. These are normalized units where the circular velocity is one at a cylindrical radius of one and height zero. Positions are therefore scaled to the physical radius that is defined to be "1" and velocities are scaled to the physical velocity at that radius. For example, for the Milky Way a reasonable choice is to normalize positions by 8 kpc and velocities by 220 km s$^{-1}$ in a model where the Sun is assumed to be at 8 kpc from the Galactic center and the circular velocity at the Sun is 220 km s$^{-1}$ (e.g., Bovy et al. 2012). Potential instances can easily be set up in this system of units by using the normalize= keyword when initializing (see below).

Functions to translate between physical units and the natural units defined above are contained in the galpy.util.bovy_conversion module. This module only contains non-trivial conversion; conversions of positions, velocities, energies, and actions are straightforward and therefore not included. Some of the conversion functions are illustrated in Figure 2. Further conversions to alternative units are also included: For example, force_in_10m13kms2 to convert forces to $10^{-13}$ km s$^{-2}$; dens_in_gevcc to convert densities to Gev cm$^{-3}$ (useful for dark-matter detection work); and dens_in_meanmatterdens to express densities in units of the mean-matter density in the Universe (useful for dealing with virial quantities).

As mentioned before, galpy will work in different units than natural units as well, as long as the system of units is consistent (e.g., the forces have to be in units consistent with the positions and velocities for the orbit integration to work). Natural units are only really depended on in situations where numerical determinations of zeros of a function or of integrals are difficult and approximations are made. An example is the evaluation of actions and frequencies, which often involve integrals that can

```
1 from galpy.potential import DoubleExponentialDiskPotential
2 dp= DoubleExponentialDiskPotential(normalize=1.,
3                                    hr=3./8.,hz=0.3/8.)
4 dp(1.,0.1) # The potential itself at R=1., z=0.1
5 -1.1037196286636572
6 dp.Rforce(1.,0.1) # The radial force
7 -0.9147659436328015
8 dp.zforce(1.,0.1) # The vertical force
9 -0.50056789703079607
10 dp.R2deriv(1.,0.1) # The second radial derivative
11 -1.0189440730205248
12 dp.z2deriv(1.,0.1) # The second vertical derivative
13 1.0648350937842703
14 dp.Rzderiv(1.,0.1) # The mixed radial,vertical derivative
15 -1.1872449759212851
16 dp.dens(1.,0.1) # The density
17 0.076502355610946121
18 dp.dens(1.,0.1,forcepoisson=True) # Using Poisson's eqn.
19 0.076446652249682681
20 dp.mass(1.,0.1) # The mass to R=1 and up to |z| = 0.1
21 0.7281629803939751
22 dp.vcirc(1.) # The circular velocity at R=1.
23 1.0 # By definition, because of normalize=1.
24 dp.omegac(1.) # The rotational frequency
25 1.0 # Also because of normalize=1.
26 dp.epifreq(1.) # The epicycle frequency
27 1.3301123099210266
28 dp.verticalfreq(1.) # The vertical frequency
29 3.7510872575640293
30 dp.flattening(1.,0.1) #The flattening (see caption)
31 0.42748757564198159
32 dp.lindbladR(1.75,m='corotation') # co-rotation resonance
33 0.540985051273488           # radius
```

Fig. 3.— Methods of Potential instances in galpy, illustrated using a double-exponential disk potential with $\ln \rho(R, z) = -R/h_r - |z|/h_z +$ constant, normalized to have a circular velocity of 1 at $R = 1$. All of the outputs are in natural units (see § 2.2). The flattening is defined as $q = \sqrt{|z F_R/(R F_z)|}$.

be difficult to evaluate for close-to-circular or very eccentric orbits. In the code, approximations are made that depend in some cases on quantities being expressed in natural units. Even if most galpy functions work well in any system of units, because galpy is being developed and tested primarily in natural units, it is recommended to use these units when using galpy.

## 3. galpy's POTENTIAL FRAMEWORK

At the heart of galpy are Potential objects. These come in three flavors: one dimensional, two dimensional, and three dimensional. The one-dimensional Potential classes only have limited use. Their main application is to serve as the vertical potential at a given location in the disk of a galaxy when orbits in the disk are approximated as separating into planar and vertical motions (see the discussion of the adiabatic approximation in § 5.3 below). Because they are not more generally useful, we will not discuss them further here.

Two-dimensional Potential classes are designed to model the potential in the plane of a disk galaxy. For example, all of galpy's current non-axisymmetric potentials are two-dimensional models. These models can be used in exactly the same way as three dimensional Potential classes, except for methods that are specific to three-dimensional models. We will focus the discussion in this section on the full, three-dimensional Potential class and instances thereof.

## 3.1. *General framework*

All three-dimensional Potential classes inherit from the general Potential class. (Two-

```
 1  from galpy.potential import Potential
 2
 3  def smoothInterp(t,dt,tform):
 4      """Smooth interpolation in time, following Dehnen (2000)"""
 5      if t < tform: smooth= 0.
 6      elif t > (tform+dt): smooth= 1.
 7      else:
 8          xi= 2.*(t-tform)/dt-1.
 9          smooth= (3./16.*xi**5.-5./8*xi**3.+15./16.*xi+.5)
10      return smooth
11
12  class TimeInterpPotential(Potential):
13      """Potential that smoothly interpolates in time between two static potentials"""
14      def __init__(self,pot1,pot2,dt=100.,tform=50.):
15          """pot1= potential for t < tform, pot2= potential for t > tform+dt, dt: time over which to turn on pot2,
16          tform: time at which the interpolation is switched on"""
17          Potential.__init__(self,amp=1.)
18          self._pot1= pot1
19          self._pot2= pot2
20          self._tform= tform
21          self._dt= dt
22          return None
23
24      def _Rforce(self,R,z,phi=0.,t=0.):
25          smooth= smoothInterp(t,self._dt,self._tform)
26          return (1.-smooth)*self._pot1.Rforce(R,z)+smooth*self._pot2.Rforce(R,z)
27
28      def _zforce(self,R,z,phi=0.,t=0.):
29          smooth= smoothInterp(t,self._dt,self._tform)
30          return (1.-smooth)*self._pot1.zforce(R,z)+smooth*self._pot2.zforce(R,z)
```

FIG. 4.— Example of a new `Potential` class: a potential that smoothly interpolates between two given potentials over time. We will use this potential in the example in Figure 21 in § 5 to investigate adiabatic changes to orbits.

dimensional `Potential` classes inherit from a general `planarPotential` class and one-dimensional classes inherit from a general `linearPotential` class). Typically, library-use of any `Potential` instance is through methods that are defined in the general `Potential` class. For basic functionality such as evaluating the potential or its forces at a point, this general `Potential` routine multiplies the subclass routines with an *amplitude* parameter; for more complicated routines such as calculating the circular velocity or the location of Lindblad frequencies, more work is done by the general class. The philosophy behind this implementation is that it allows users to implement additional potentials with a minimum of effort, only requiring the basic properties of a potential to be implemented. Allowing the general `Potential` class to multiply in an amplitude is helpful for using the natural units described above.

A basic implementation of a new specific potential in `galpy` requires the user to write a class that inherits from the general `galpy.potential.Potential` class and that has methods `_evaluate`, `_Rforce`, and `_zforce`. Any new class should take an `amp` parameter upon initialization and pass this to the general `Potential` class' initializer, by calling `galpy.potential.Potential.__init__(self,amp=amp)` within the new class' `__init__` function. More advanced functionality or non-axisymmetric potentials require additional functions to be implemented specifying the second derivatives and derivatives with respect to azimuth. All of the pure-`python` functionality of `galpy` can then be applied to the new potential. While the forces and second derivatives could be obtained by taking numerical derivatives of the potential, such that a new potential could be defined using only the `_evaluate` function, this is not currently supported. The test suite described in § 8.2 automatically checks for new poten-tials defined at the top-level of `galpy` that the forces are the negative derivative of the potential and that the second derivatives are correct, by using numerical derivatives. The density can be explicitly implemented using the `_dens` method; if it is not included then the density is computed by using the Poisson equation if all of the relevant second derivatives are implemented.

Many of the methods available for any `Potential` instance are illustrated in Figure 3, using a double-exponential disk potential as an example. In this example, we use the `normalize=1.` keyword to normalize the potential such that it has a circular velocity of 1 at $R = 1$ and $z = 0$, the preferred set of units in `galpy` (see § 2.2). The `normalize=X` keyword in general normalizes the potential such that the radial force at $R = 1$ is equal to $X$. This can be used to normalize a sum of potentials to have a circular velocity of 1 at $R = 1$, by making sure that the individual potentials $i$'s $X_i$ sum to 1.

I give an example of a new `Potential` class in Figure 4. This class implements a potential that smoothly interpolates between two given potentials as a function of time. We will use it in § 5 to look at changes to orbits when the potential is changed adiabatically. Once this new `Potential` class is defined, it can be used like any other potential (note that use of some methods requires the second derivatives of the potential to be implemented as well).

Almost all functions in `galpy` that take a `Potential` instance as an argument can also take lists of these. That is, new potentials can be specified using arbitrary combinations of basic potentials. When the potential is used, the contribution from each constituent potential is added to the total potential, force, etc. The `galpy.potential` module contains functions `evaluatePotentials`, `evaluateRforces`, `evaluatephiforces`, `evaluatezforces`, as well as sec-

C code:

```
1   # LogarithmicHaloPotential.c
2   #include <math.h>
3   #include <galpy_potentials.h>
4   //3 arguments: amp, c2, and q
5   double LogarithmicHaloPotentialEval(double R,double Z, double phi,
6                                       double t,
7                                       struct potentialArg * potentialArgs){
8       double * args= potentialArgs->args;
9       double amp= *args;
10      double q= *(args+1);
11      double c2= *(args+2);
12      double zq= Z/q;
13      return 0.5 * amp * log(R*R+zq*zq+c2);
14  }
15  double LogarithmicHaloPotentialRforce(double R,double Z, double phi,
16                                        double t,
17                                        struct potentialArg * potentialArgs){
18      double * args= potentialArgs->args;
19      double amp= *args++;
20      double q= *args++;
21      double c2= *args--;
22      double zq= Z/q;
23      return - amp * R/(R*R+zq*zq+c2);
24  }
25  double LogarithmicHaloPotentialzforce(double R,double z,double phi,
26                                        double t,
27                                        struct potentialArg * potentialArgs){
28      double * args= potentialArgs->args;
29      double amp= *args++;
30      double q= *args++;
31      double c2= *args--;
32      double zq= z/q;
33      return -amp * z/q/q/(R*R+zq*zq+c2);
34  }
35  ...
36  # In galpy/potential_src/potential_c_ext/galpy_potentials.h
37  ...
38  //LogarithmicHaloPotential
39  double LogarithmicHaloPotentialEval(double ,double , double, double,
40                                      struct potentialArg *);
41  double LogarithmicHaloPotentialRforce(double ,double , double, double,
42                                        struct potentialArg *);
43  double LogarithmicHaloPotentialzforce(double,double,double,double,
44                                        struct potentialArg *);
45  ...
46  # In galpy/orbit_src/orbit_c_ext/integrateFullOrbit.c's parse_leapFuncArgs_Full
47  ...
48      case 0: //LogarithmicHaloPotential, 2 arguments, unique case integer should match the integer in line 64 below
49          potentialArgs->Rforce= &LogarithmicHaloPotentialRforce;
50          potentialArgs->zforce= &LogarithmicHaloPotentialzforce;
51          potentialArgs->phiforce= &ZeroForce;
52          potentialArgs->nargs= 3;
53          break;
54      ...
```

python code:

```
61  # In galpy/orbit_src/integrateFullOrbit.py's _parse_pot
62  ...
63      if isinstance(p,potential.LogarithmicHaloPotential):
64          pot_type.append(0) # See line 48 above
65          pot_args.extend([p._amp,p._q,p._core2])
66  ...
67  # In galpy/potential_src/LogarithmicHaloPotential.py's __init__
68  ...
69          self.hasC= True
```

FIG. 5.— Minimal **C** implementation of **galpy**'s **LogarithmicHaloPotential**. This figure illustrates the steps necessary for implementing a potential in C in such a way that it can automatically be used in the **galpy** framework. The top file is a new **C** file that has the implementation of the potential and its radial and vertical force law. These new **C** functions need to be declared in the general **galpy_potentials.h** header file on line 42. The following two files that require editing provide the glue between **C** and **python**: **integrateFullOrbit.c**'s **parse_leapFuncArgs_Full** function contains the code that specifies the functions that implement the forces and the number of parameters (line 52); the **python** function **_parse_pot** in **integrateFullOrbit.py** contains the code that stores the potential parameters in an array that will be passed to the **C** code (line 61). Finally, the initialization of the **LogarithmicHaloPotential** instance needs to set the attribute **hasC** to **True**, such that all relevant code can automatically use **C** for speeding up calculations. To use this potential for action–angle calculations in **C**, more glue needs to be written that is similar to that in **integrateFullOrbit.c**.

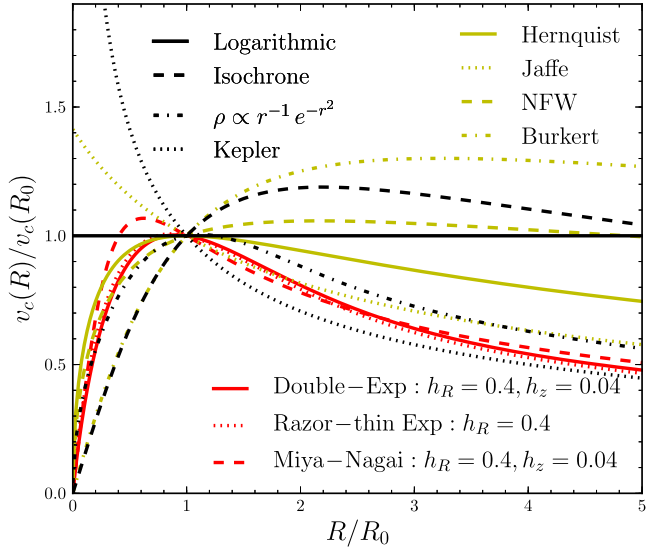

FIG. 6.— Rotation curves of some of the axisymmetric potentials included in `galpy`. All rotation curves are in *natural units* where the circular velocity $v_c = 1$ at $R = 1$. Unless otherwise indicated, each potential's scale parameter is set to one. This figure was made by calling each `Potential` instance's `plotRotcurve` method.

ond derivatives `evaluateR2derivs`, `evaluatez2derivs`, `evaluateRzderivs`, and `evaluateDensities` that can be used to evaluate `Potential` functions for lists of potentials. Similar functions exist for the other `Potential` methods used in Figure 3 (e.g., `vcirc`). New functionality that makes use of `Potential` objects should be written using these functions to evaluate the potential and its properties to sustain the support for lists of `Potential` objects. That is, one should make use of functions that can act on lists of `Potential` instances, such as `evaluateRforces`, rather than instance methods, such as `Rforce`.

### 3.2. *C implementations of potential classes*

In certain parts of the codebase, `galpy` uses C to speed up computations. Currently, this is limited to orbit integration and the calculation of certain types of action–angle coordinates. To make use of this functionality, `Potential` classes implemented in `python` need to also be explicitly implemented in C by the user and this needs to be registered by setting the `hasC` attribute of the `python` `Potential` subclass in question to `True`. This procedure is illustrated in Figure 5 in the case of `LogarithmicHaloPotential`, a simple logarithmic potential contained in `galpy`. This minimal implementation only implements the potential itself and the force components; glue is written to make it possible to call this function from within `python`. By following the procedure given in Figure 5, new, user-contributed potentials will be automatically incorporated into the `galpy` framework: wherever the code automatically switches to C routines to speed up the code, this will be done for the new `Potential` subclass as well, whether it is used on its own or whether it is contained in a list of `Potential` instances, all of which have their own C implementation.

Full details of the procedure for adding C implementations can be found in the online documentation at

http://galpy.readthedocs.org/en/latest/potential.html#adding-potentials-to-the-galpy-framework .

All potentials in `galpy` should have `python` implementations as many functions only use the `python` `Potential` methods; the C implementations are only an additional feature for potentials to allow certain computations to be sped up.

### 3.3. *Axisymmetric potentials*

`galpy` contains a large number of axisymmetric, three-dimensional potentials that can be combined to form a realistic model for a galaxy. Many of the methods that are defined for each potential are illustrated in Figure 3. The rotation curves for a subset of the axisymmetric potentials are shown in Figure 6. Some of the potentials shown in this figure are implemented as special cases of more general potential classes. This is the case in particular for the `KeplerPotential`, which is a special case of general `PowerSphericalPotentials` with densities $\rho \propto r^{-\alpha}$; the $\rho \propto r^{-1} e^{-r^2}$, which is a special case of the `PowerSphericalPotentialwCutoff` potential with density $\rho \propto r^{-\alpha} \exp\left(-(r/rc)^2\right)$; and the triplet of Hernquist, Jaffe, and NFW potentials, which are subclasses of a general `TwoPowerSphericalPotential` with density $\rho \propto (r/a)^{-\alpha} (1 + r/a)^{\alpha-\beta}$.

Because some of the axisymmetric potentials' evaluation of the potential itself or of the derivatives is computationally expensive, a general class `interpRZPotential` is provided that tabulates the potential and its forces on a grid for a given axisymmetric potential and uses spline interpolation to evaluate the potential. This is implemented simply by interpolating separately the potential itself, the forces and second derivatives, and the circular velocity curve, its derivative, and the various frequencies (epicycle and vertical). That is, no effort is made to calculate the forces using derivatives of the splines interpolating the potentials, etc. Therefore care should be taken to build a dense enough interpolation grid to avoid large inconsistencies between the potential and its forces and derivatives.

Interpolated potentials implemented through `interpRZPotential` can be used wherever `Potential` instances can be used anywhere in `galpy`. This includes functions that use C to speed up computations; to use the latter feature one needs to specify `enable_c=True` when setting up the `interpRZPotential` instance, which stores the potential and force tabulations in a way in which it can be used in C and sets the `hasC` attribute (see § 3.2) to `True` to register that the instance has a C interface.

### 3.4. *Non-axisymmetric potentials*

`galpy` contains a number of non-axisymmetric potentials that can be used to investigate the effect of non-axisymmetry on the dynamical structure of galaxies. Currently, all of these are two-dimensional potential with a single exception. Primarily, these potentials represent simple parametric forms to represent perturbations to the potential of disk galaxies. `galpy` contains a general $\cos m\phi$ perturbation as `CosmphiDiskPotential` with potential $\Phi(R,\phi) \propto R^p \cos(m(\phi - \phi_b))$. The special cases corresponding to a lopsided disk (`LopsidedDiskPotential`; $m = 1$) and an elliptical disk (`EllipticalDiskPotential`; $m = 2$; Kuijken & Tremaine 1994) are also included; these two are also implemented in C, while the more general version currently is not. These models can all be turned on smoothly in time by specifying a time `tform` when the perturbation starts growing and a time period `tsteady` over which it is fully grown; the growth function is that of Dehnen (2000).

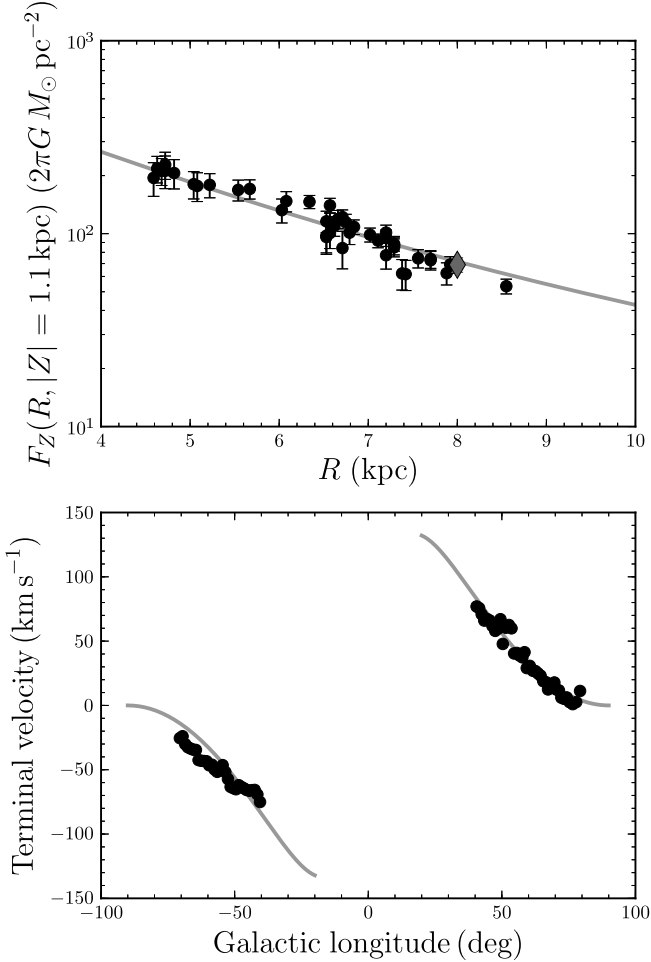

FIG. 7.— Radial properties of MWPotential2014. The top panel shows the radial profile of the vertical force at 1.1 kpc above the mid-plane and compares it to the measurements of Bovy et al. (2013) (black points) and of Zhang et al. (2013) (gray diamond). The bottom panel displays the terminal-velocity curve of MWPotential2014 and compares it to the data from Clemens (1985) and McClure-Griffiths & Dickey (2007).

`galpy` also contains models for the time-dependent perturbations coming from spiral structure and the bar. For spiral structure this is a simple logarithmic spiral model $\Phi(R, \phi, t) \propto \cos(\alpha \ln R - m(\phi - \Omega_s t - \gamma))$ that can either be steady (`SteadyLogSpiralPotential`) or transient (`TransientLogSpiralPotential`). The steady model can be turned on slowly in a similar way as the $\cos m\phi$ perturbations above, to allow for adiabatic growth of the spiral. The transient spiral has an amplitude $\propto \exp(-[t - t_0]^2/2\sigma^2)$. The bar potential that is contained in `galpy` is the simple rotating quadrupole of Dehnen (2000): `DehnenBarPotential`.

The one non-axisymmetric potential that is not limited to two dimensions is the potential corresponding to a moving object, `MovingObjectPotential`. This potential is initialized by giving an integrated `Orbit` instance (see below) and a mass. This potential can then be added to the potential that the moving object is integrated in to include the gravity from the moving object on test particles. This can be used, for example, to simulate the impact of molecular clouds on stellar orbits in the disk.

### 3.5. *Example: Milky-Way-like potentials*

The `galpy.potential` module also contains a model `MWPotential2014` for the Milky Way's gravitational potential that is designed to provide a simple, easy-to-use model in cases where a realistic model for the Milky Way is required. This model is fit to some of the dynamical data on the Milky Way as described in this section. However, it is important to keep in mind that this model is merely provided for convenience and that it is not designed to provide the best possible current model for the Milky Way's gravitational forces. Any application that is significantly affected by the current uncertainty in the potential should use the various tools in the `galpy` library to explore the dependence on different Milky Way potential parameters and functional forms.

We perform a fit that is very similar to that described in § 5 of Bovy et al. (2013), but uses some additional data to provide a more realistic description of the Milky Way on small and large scales. The potential model is a simplified version of that in Bovy et al. (2013) and consists of a bulge modeled as a power-law density profile that is exponentially cut-

TABLE 1
PARAMETERS AND PROPERTIES OF MWPotential2014

| Parameter | MWPotential2014 | Constraint |
|---|---|---|
| $R_0$ (kpc) | 8 | fixed |
| $v_c(R_0)$ (km s$^{-1}$) | 220 | fixed |
| $f_b$ | 0.05 | ... |
| $f_d$ | 0.60 | ... |
| $f_h$ | 0.35 | ... |
| Bulge power−law exponent | −1.8 | fixed |
| Bulge cut−off radius (kpc) | 1.9 | fixed |
| $a$ (kpc) | 3.0 | ... |
| $b$ (pc) | 280 | ... |
| Halo $r_s$ (kpc) | 16 | ... |
| | | |
| $\sigma_b$ (km s$^{-1}$) | 109 | $117 \pm 15$ |
| $F_Z(R_0, 1.1 \text{ kpc})$ $(2\pi G \, M_\odot \text{ pc}^{-2})$ | 72 | $67 \pm 6$ |
| $\Sigma_{vis}(R_0)$ $(M_\odot \text{ pc}^{-2})$ | 53 | $55 \pm 5$ |
| $F_Z$ scale length (kpc) | 3.2 | $2.7 \pm 0.1$ |
| $\rho(R_0, z=0)$ $(M_\odot \text{ pc}^{-3})$ | 0.10 | $0.10 \pm 0.01$ |
| $(\mathrm{d} \ln v_c/\mathrm{d} \ln R)\|_{R_0}$ | −0.10 | −0.2 to 0 |
| $M(r < 60 \text{ kpc})$ $(10^{11} M_\odot)$ | 4.1 | $4.0 \pm 0.7$ |
| | | |
| $M_b$ $(10^{10} M_\odot)$ | 0.5 | ... |
| $M_d$ $(10^{10} M_\odot)$ | 6.8 | ... |
| $R_d$ (kpc) | 2.6 | ... |
| $\rho_{DM}(R_0)$ $(M_\odot \text{ pc}^{-3})$ | 0.008 | ... |
| $M_{vir}$ $(10^{12} M_\odot)$ | 0.8 | ... |
| $r_{vir}$ (kpc) | 245 | ... |
| Concentration | 15.3 | ... |
| $v_{esc}(R_0)$ (km s$^{-1}$) | 513 | ... |

NOTE. — The top part of this table gives the explicit parameters of the MWPotential2014 model: the relative contribution from the bulge ($f_b$), disk ($f_d$), and halo ($f_h$) to the radial force at $R_0$; the power-law exponent and the exponential cut-off radius of the bulge density; the scale length $a$ and scale height $b$ of the Miyamoto-Nagai disk model; and the scale radius $r_s$ of the NFW halo model. The second section displays MWPotential2014's values for some of the constraints that the parameters of the model are fit to. The final part of this table gives some additional properties of the potential.

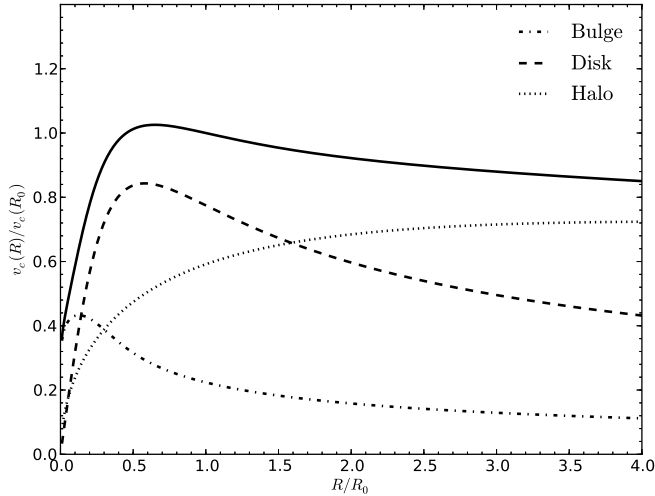

FIG. 8.— The rotation curve of `MWPotential2014` out to 32 kpc and its decomposition into bulge, disk, and halo contributions.

off (`PowerSphericalPotentialwCutoff`) with a power-law exponent of $-1.8$ and a cut-off radius of 1.9 kpc; a `MiyamotoNagaiPotential` disk; and a dark-matter halo described by an `NFWPotential`. The relative amplitudes of these three components, the scale length and height of the disk potential, and the scale radius of the NFW halo are fit to the following data:

a) The velocity dispersion $\sigma_b = 117 \pm 15\,\mathrm{km\,s^{-1}}$ in Baade's window (Dehnen & Binney 1998; Binney & Tremaine 2008);

b) The vertical force at the solar circle at 1.1 kpc from the plane $|F_Z| = 67 \pm 6\,(2\pi G\,M_\odot\,\mathrm{pc^{-2}})$ and the local visible surface density $\Sigma = 55 \pm 5\,M_\odot\,\mathrm{pc^{-2}}$ from Zhang et al. (2013);

c) The vertical force measurements at 1.1 kpc from the plane of Bovy et al. (2013);

d) The terminal-velocity measurements of Clemens (1985) and McClure-Griffiths & Dickey (2007), modeled in the same way as in Bovy et al. (2013);

e) The measurement of the mid-plane density at the solar circle of Holmberg & Flynn (2000): $\rho(R_0, Z = 0) = 0.10 \pm 0.01\,M_\odot\,\mathrm{pc^{-3}}$;

f) The constraint on the logarithmic slope of the rotation curve from Bovy et al. (2012), represented in the same way as in equation (41) in Bovy et al. (2013);

g) The measurement of the total mass within 60 kpc from Xue et al. (2008): $M(r < 60\,\mathrm{kpc}) = 4.0 \pm 0.7 \times 10^{11}\,M_\odot$.

In addition to these constraints, we set the solar distance to the Galactic center to $R_0 = 8$ kpc and the circular velocity at the Sun to $V_0 = 220\,\mathrm{km\,s^{-1}}$ (Bovy et al. 2012). We fit the potential model to these data and then adjust the parameters to a close, simple number. The parameters of the resulting `MWPotential2014` are shown in Table 1.

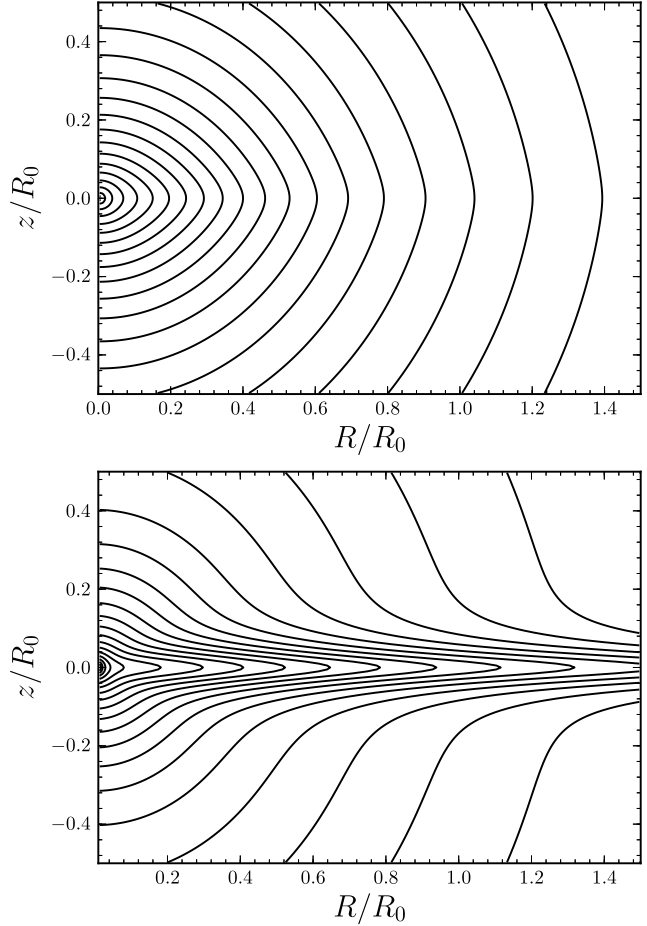

FIG. 9.— Contours of the potential and density profile of `MWPotential2014` in the region of the disk. Potential contours are linearly spaced and density contours use logarithmic spacing.

We compare the properties of `MWPotential2014` to the data that it was fit to in the second part of Table 1 and in Figure 7. The "$F_Z$ scale length" that is shown in Table 1 is an approximate exponential scale length of the vertical force at 1.1 kpc between 4 and 9 kpc, which is compared to the scale length obtained from an exponential fit to the data of Bovy et al. (2013), which are also displayed in Figure 7.

Other properties of `MWPotential2014` are shown in Figures 8 and 9 and in the third part of Table 1. Figures 8 displays the rotation curve and its decomposition into bulge, disk, and dark-halo contributions. The masses of the three components are given in Table 1; the virial properties (radius, mass, and concentration) of the NFW halo are also presented there. $R_d$ is an approximate exponential scale length from a fit to the Miyamoto-Nagai-disk density, which is compared to the determination of $R_d = 2.15 \pm 0.14$ kpc of Bovy et al. (2013). The local dark-matter density $\rho_{\mathrm{DM}} = 0.008\,M_\odot\,\mathrm{pc^{-3}}$ is in good agreement with current constraints (e.g., Bovy & Tremaine 2012; Zhang et al. 2013; Piffl et al. 2014) as is the escape velocity (e.g., Smith et al. 2007). Figure 9 shows contours of the potential and the density of `MWPotential2014` in the region close to the disk.

`MWPotential2014` does not contain the supermassive black hole at the center of the Milky Way. If the

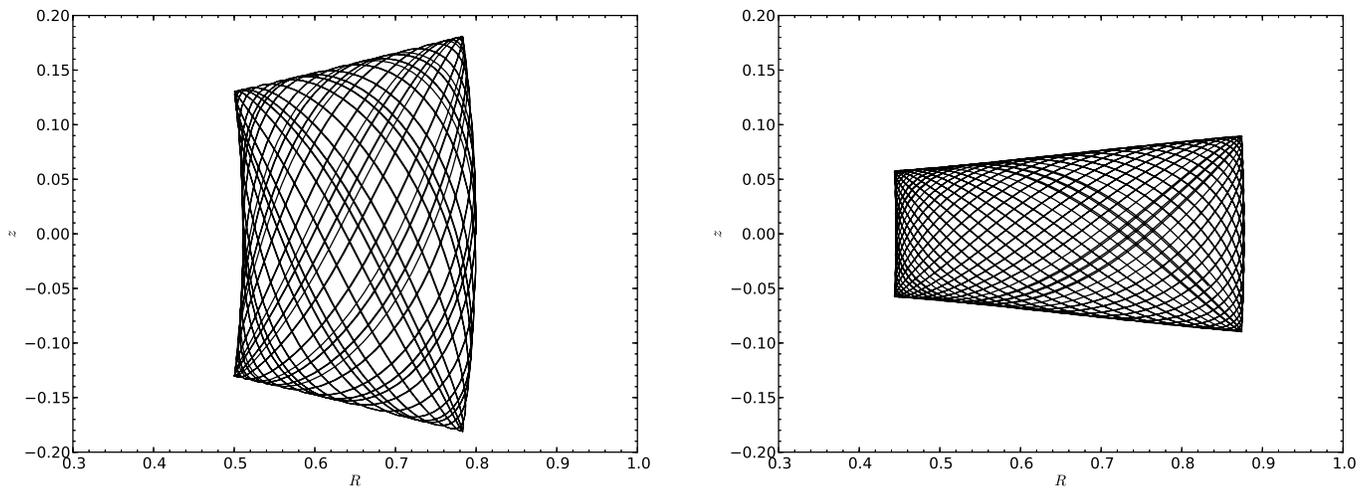

```
1  import numpy
2  from galpy.potential import MWPotential2014
3  from galpy.potential import evaluatePotentials as evalPot
4  from galpy.orbit import Orbit
5  E, Lz= -1.25, 0.6
6  o1= Orbit([0.8,0.,Lz/0.8,0.,numpy.sqrt(2.*(E-evalPot(0.8,0.,MWPotential2014)-(Lz/0.8)**2./2.)),0.])
7  ts= numpy.linspace(0.,100.,2001)
8  o1.integrate(ts,MWPotential2014)
9  o1.plot(xrange=[0.3,1.],yrange=[-0.2,0.2],color='k')
10 o2= Orbit([0.8,0.3,Lz/0.8,0.,numpy.sqrt(2.*(E-evalPot(0.8,0.,MWPotential2014)-(Lz/0.8)**2./2.-0.3**2./2.)),0.])
11 o2.integrate(ts,MWPotential2014)
12 o2.plot(xrange=[0.3,1.],yrange=[-0.2,0.2],color='k')
```

FIG. 10.— Orbit integrations of two orbits with the same energy and angular momentum in `MWPotential2014`. The two orbits are displayed in the meridional plane at the top and the code used to generate them is shown at the bottom.

black hole's gravity needs to be included, e.g., for tracing the orbits of stars or pulsars kicked out from the Galactic center (e.g., Dexter & O'Leary 2013), a `KeplerPotential` can be added with a mass of $4 \times 10^6 \, M_\odot$ (e.g., Gillessen et al. 2009) as

```
MWPotential2014.append(KeplerPotential(amp=\
    4*10**6./bovy_conversion.mass_in_msol(220.,8.)))
```

Finally, note that `galpy` contains an older version of a similar potential: `MWPotential`. This model was *not* fit to data on the Milky Way (although it is close to a good model) and is now superseded by `MWPotential2014`.

## 4. ORBIT INTEGRATION

### 4.1. *General framework*

The integration and characterization of orbits is an essential part of `galpy`. Orbit integration is supported through a number of different integrators (see § 4.2 below) and for phase–space dimensionalities from two through six. These five flavors of orbits are (a) `linearOrbit` for the integration of orbits in one-dimensional potentials, (b) `planarOrbit` and `planarROrbit` for orbits in non-axisymmetric and axisymmetric two-dimensional potentials, and (c) `FullOrbit` and `RZOrbit` for three-dimensional orbits. The `linearOrbits` are useful for investigating motions perpendicular to the mid-plane of a disk galaxy, when these motions are assumed to decouple from the motions in the plane. The axisymmetric two- and three-dimensional orbits `planarROrbit` and `RZOrbit` assume conservation of angular momentum and do not keep track of the azimuthal angle. `planarOrbit` and `FullOrbit` do keep track of the azimuthal angle and integrate the equations of motions without assuming any symmetry.

While the different flavors of orbits are all implemented as different classes deriving from a superclass `OrbitTop`, this structure is entirely hidden from the user and all public interfacing with instances of these classes is through a general `Orbit` class. All types of orbits are instantiated through this class by providing the initial conditions of the orbit and the type of orbit is determined from the dimensionality of this initial condition. Typically the initial condition is specified by an array of positions and velocities in the cylindrical Galactocentric coordinate frame $(R, v_R, v_T, z, v_z, \phi)$ or lower-dimensional projections of this: $(x, v_x)$ for `linearOrbit` instances; $(R, v_R, v_T[, \phi])$ for `planarOrbit` and `planarROrbit` objects (the latter do not specify $\phi$); and $(R, v_R, v_T, z, v_z[, \phi])$ for `FullOrbit` and `RZOrbit` objects. For data analysis in the Milky Way, `Orbit` instances can also be specified in close-to-observable coordinates: $(\alpha, \delta, D, \mu_{\alpha,*}, \mu_\delta, v_{\mathrm{los}})$ or RA, Dec, distance, proper motions in $(\alpha, \delta)$, and line-of-sight velocity (where $\mu_{\alpha,*} = \mu_\alpha \cos(\delta)$); initial conditions may also similarly be specified using Galactic coordinates and velocities can be given as $(U, V, W)$. To use this functionality, the coordinate transformation between heliocentric and Galactocentric coordinates needs to be specified by the user by giving the Sun's distance to the Galactic center $R_0$ and the mid-plane $z_0$ and the Sun's velocity with respect to the Galactic center. The latter is split into the Sun's motion with respect to the circular velocity $v_c(R_0)$ and

```
1  from galpy.orbit import Orbit
2  from galpy.potential import MWPotential2014
3  o= Orbit([0.8,0.3,0.75,0.,0.2,0.])  # setup R,vR,vT,z,vz,phi
4  times= numpy.linspace(0.,10.,1001)  # Output times
5  o.integrate(times,MWPotential2014)  # Integrate
6  o.E()  # Energy
7  -1.2547650648697966
8  o.L()  # Angular momentum
9  array([[ 0. , -0.16, 0.6 ]])
10 o.Jacobi(OmegaP=0.65)  #Jacobi integral E-OmegaP Lz
11 array([-1.64476506])
12 o.ER(times[-1]), o.Ez(times[-1])  # Rad. and vert. E at end
13 (-1.27601734263047, 0.02125220184785190 )
14 o.rperi(), o.rap(), o.zmax()  # Peri-/apocenter r, max. |z|
15 (0.44231993168097, 0.87769030382105, 0.07745325735289016)
16 o.e()  # eccentricity (rap-rperi)/(rap+rperi)
17 0.32982348199330563
18 o.R(2.,ro=8.)  # Cylindrical radius at time 2. in kpc
19 3.547077287620007
20 o.vR(5.,vo=220.)  # Cyl. rad. velocity at time 5. in km/s
21 45.202530965094553
22 o.ra(1.), o.dec(1.)  # RA and Dec at t=1. (default settings)
23 (array([ 288.19277]), array([ 18.98069155]))
24 o.jr(type='adiabatic'), o.jz()  # R/z actions (ad. approx.)
25 (0.05285302231137586, 0.0066379888500751242)
26 # Rad. period w/ Staeckel approximation w/ focal length 0.5,
27 o.Tr(type='staeckel',delta=0.5,ro=8.,vo=220.)  # in Gyr
28 0.103946786401846446
29 o.plot(d1='R',d2='z')  # Plot the orbit in (R,z)
30 ...
31 o.plot3d()  # Plot the orbit in 3D, w/ default [x,y,z]
```

FIG. 11.— Methods of `Orbit` instances, illustrated using an orbit similar to the one at the right in Figure 10. The radial energy (rad. E) is given by $E_R = \Phi(R,0) + v_R^2/2 + v_T^2/2$; the vertical energy (vert. E) is defined as $E_z = \Phi(R,z) - \Phi(R,0) + v_z^2/2$. All outputs are in natural units (see § 2.2), except for those with a physical distance or velocity scale set through `ro=` or `vo=`, respectively. The action–angle methods and types corresponding to the adiabatic (ad. approx.) and Staeckel approximations are discussed further in § 5. `Orbit` instances have many more methods similar to `o.R()`, `o.ra()`, and `o.jr()`.

$v_c(R_0)$ itself. This way $R_0$ and $v_c(R_0)$ can be used to normalize the coordinates to `galpy`'s natural coordinates (see § 2.2); this is done automatically. An example `Orbit` instantiation from observed coordinates looks like

```
from galpy.orbit import Orbit
o= Orbit([25.,10.,2.,5.,-2.,50.],radec=True,ro=8.,
         vo=220.,solarmotion=[-11.1,25.,7.25])
```

where the solar motion is such that the Sun's rotational velocity with respect to Sgr A* is 245 km s$^{-1}$ (Bovy et al. 2012).

The initial conditions can be integrated in time in any of the potentials by calling the `integrate` method. Figure 10 demonstrates orbit integration in `galpy`: the two orbits displayed in the top panel of this figure have the same energy and angular momentum and are integrated in the `MWPotential2014` potential; the `python` code that produces the top panels is given in the bottom panel as an illustration of the use of `galpy`. This figure is similar to figure 3.4 in Binney & Tremaine (2008).

After orbit integration, the orbit's characteristics and time dependence can be accessed through a variety of instance methods, which are illustrated in Figure 11. One property of `Orbit` instance methods is that they can produce output in physical coordinates rather than natural coordinates by specifying a distance and velocity scale through `ro=` and `vo=`, respectively. If these scales are set as keywords during the `Orbit` initialization, they are automatically used for any output; if not they can be

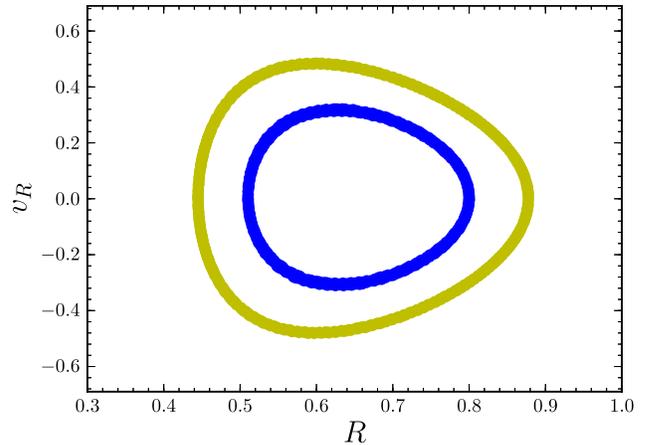

```
1  def surface_section(Rs,zs,vRs):
2    # Find points where the orbit crosses z from - to +
3    shiftzs= numpy.roll(zs,-1)
4    indx= (zs[:-1] < 0.)*(shiftzs[:-1] > 0.)
5    return (Rs[:-1][indx],vRs[:-1][indx])
6  # Calculate and plot the surface of section
7  ts= numpy.linspace(0.,1000.,20001)  # long integration
8  o1.integrate(ts,MWPotential2014)
9  o2.integrate(ts,MWPotential2014)
10 sect1Rs,sect1vRs=surface_section(o1.R(ts),o1.z(ts),o1.vR(ts))
11 sect2Rs,sect2vRs=surface_section(o2.R(ts),o2.z(ts),o2.vR(ts))
12 from matplotlib.pyplot import plot, xlim, ylim
13 plot(sect1Rs,sect1vRs,'bo',mec='none')
14 xlim(0.3,1.); ylim(-0.69,0.69)
15 plot(sect2Rs,sect2vRs,'yo',mec='none')
```

FIG. 12.— Poincaré section $(R, v_R, z = 0, v_z > 0)$ of the two orbits displayed in Figure 10 (top panel). The left orbit from Figure 10 is shown in blue and the right orbit is displayed in yellow. The code used to generate this surface of section is given in the bottom panel; this code requires the code displayed at the bottom of Figure 10 to be run first.

specified for each individual method, which also overrides the values set at initialization. This behavior can be turned off by calling the `turn_physical_off()` method; this is necessary in particular for orbits initialized from observed coordinates, as these always require `ro=` and `vo=` to be set when initializing.

Another example of how `galpy`'s orbit routines can be used is shown in Figure 12, where the Poincaré section of the two orbits displayed in Figure 10 is computed to demonstrate that these orbits have a third integral of motion in addition to the energy and angular momentum. Calculating Poincaré sections is not currently supported by `galpy`, but the code in the bottom panel of Figure 12 makes clear how simple it is to extend `galpy` to calculate other properties of orbits.

### 4.2. Supported integrators

Orbit integration in `galpy` is supported using eight different integration methods, specified using the `method=` keyword of the `integrate` method. Two of these are pure `python` based methods. The first of these is `odeint`, which corresponds to `scipy`'s ordinary-differential-equation solver of the same name. This solver uses the lsoda routine in the FORTRAN library odepack (Hindmarsh 1983). The second is `leapfrog`, which is a custom implementation of a leapfrog integrator, a second-order symplectic integrator (e.g.,

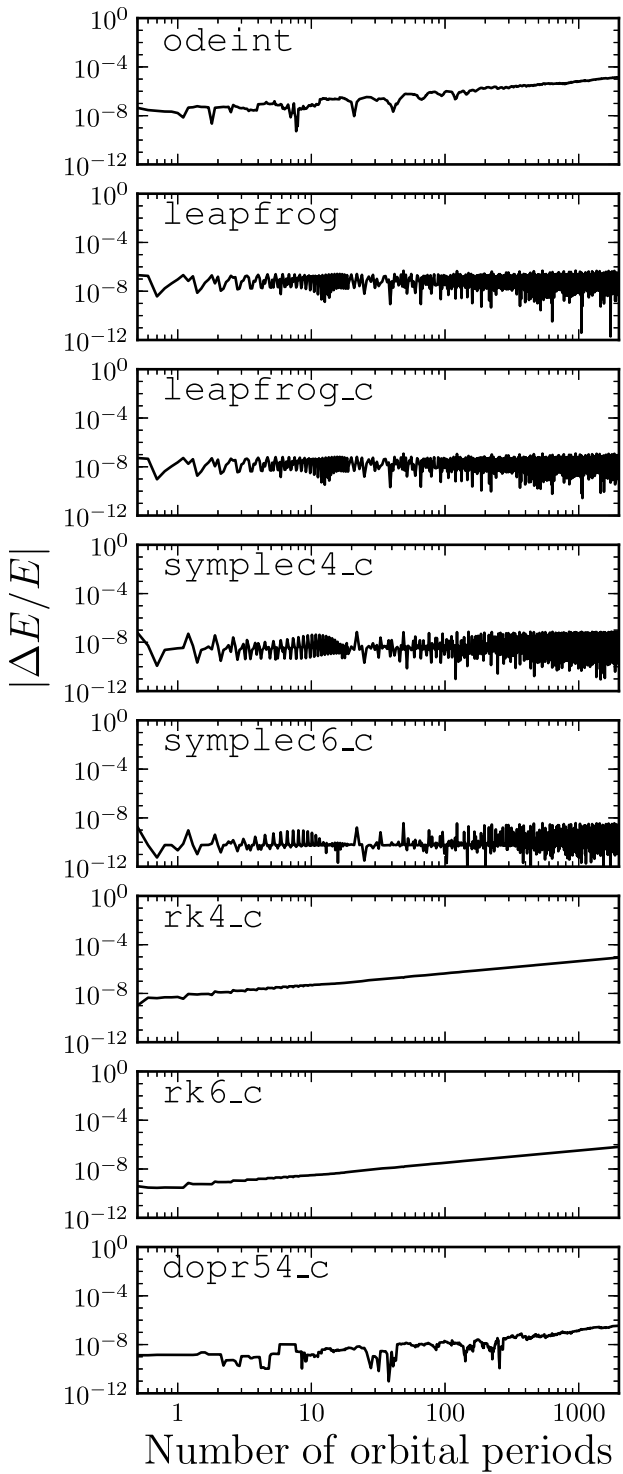

FIG. 13.— Fractional energy error as a function of the length of the integration for the orbit on the right in Figure 10 in MWPotential2014 for the eight orbit integrators contained in galpy. The top two integrators are pure python implementations: odeint is scipy's ordinary-differential-equation solver and leapfrog is a python implementation of the leapfrog integrator. The other six integrators are written in C: second (leapfrog_c), fourth (symplec4_c), and sixth (symplec6_c) order symplectic integrators; fourth, fifth, and sixth order Runge–Kutta solvers (rk4_c, dopr54_c, and rk6_c). The length of the integration is specified in units of the rotational period $T_\phi$ of the orbit. The energy error remains constant for the symplectic integrators while it increases in time for the other integrators.

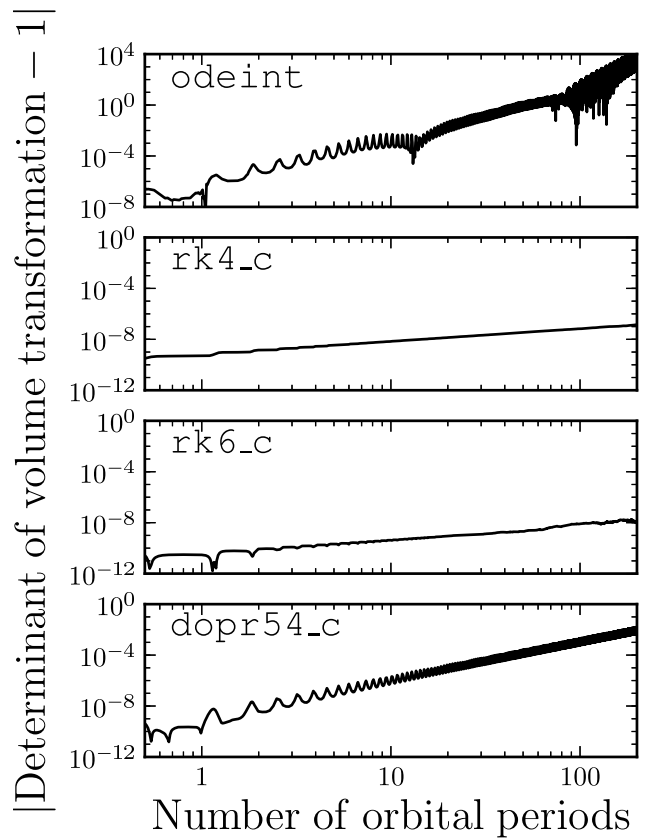

FIG. 14.— Error in the conservation of phase–space volume for the four non-symplectic integrators in galpy. The Jacobian of the volume transformation is computed by integrating $(d\mathbf{x}, d\mathbf{v})$ for an orbit with energy $-1.25$ and angular momentum $0.6$ in the mid-plane of the MWPotential2014. All integrators except for odeint use a constant time step determined at the initial condition to provide a relative and absolute error equal to or less than $10^{-8}$. The scipy integrator odeint leads to large errors in the volume integration, even for moderate numbers of orbital periods (note the different range on the $y$ axis for odeint). The rk4_c and rk6_c solvers have small volume errors, even for hundreds of orbital periods; they are also the fastest for this particular orbit.

Binney & Tremaine 2008). These integrators can be used to integrate orbits in any potential implemented in the galpy Potential framework.

For faster orbit integration, higher-order solvers are implemented in C and these can be used to integrate orbits in all galpy potentials that have C implementations (see § 3.2). Three of these integrators are Runge–Kutta solvers: the classical fourth-order Runge–Kutta method rk4_c, a fifth-order Dormand–Prince 5(4) method dopr54_c (Dormand & Prince 1980), and a sixth order method rk6_c. galpy also contains C implementations of three symplectic integrators. The first is the same as leapfrog, but coded in C: leapfrog_c. The others are a fourth-order symplectic integrator from Forest & Ruth (1990) (that corresponding to their equation [4.9]) and the SI6A sixth-order symplectic integrator from Kinoshita et al. (1991). All of the integrators except for odeint use a constant stepsize that is set at the initial condition to produce a relative and absolute error smaller than $10^{-8}$ (combining the relative and absolute error into an overall error as abs. err. + rel. err. $|(\mathbf{x}, \mathbf{v})|$). For this combination of position and velocity to be meaningful, use of galpy's natural coordinates (see

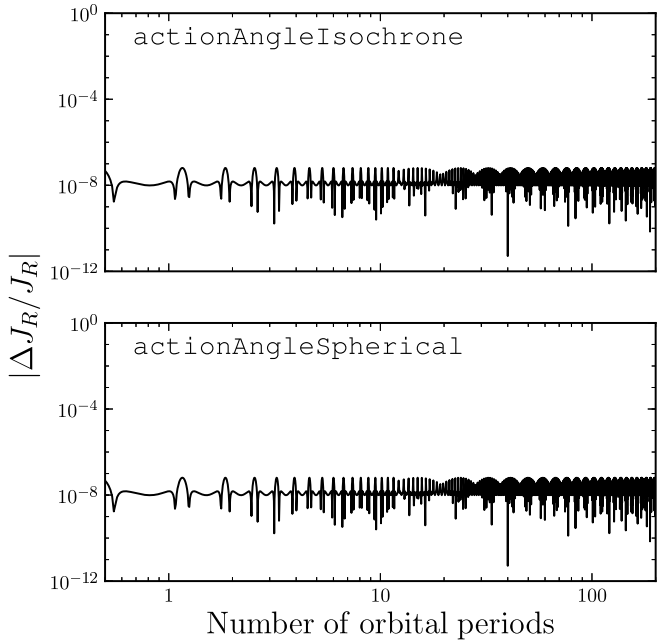

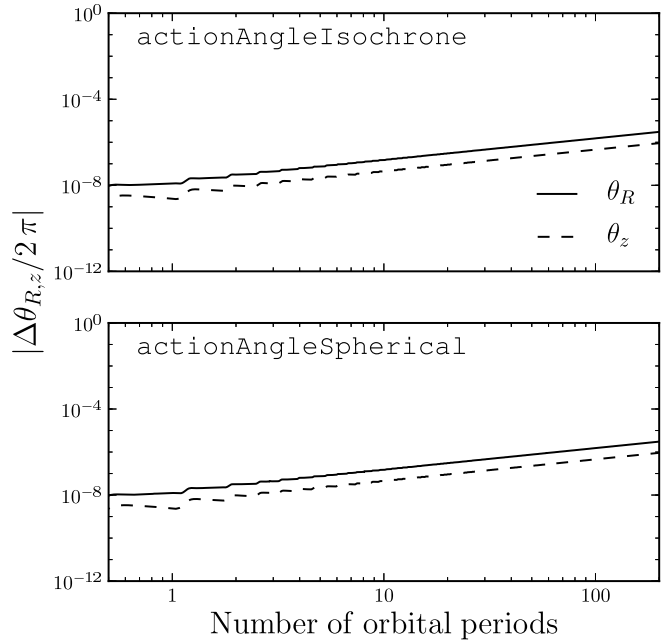

FIG. 15.— Variation of the radial action as a function of integration time for spherical potentials. The radial action is computed at each point along an orbit with the same initial condition as the orbit displayed on the right in Figure 10, but integrated in an `IsochronePotential` with scale parameter 0.3 (this isochrone potential has a similar rotation curve as `MWPotential2014` over the extent of the orbit shown in Figure 10). The radial action is computed by using the analytic formulae for an isochrone potential (Binney & Tremaine 2008) using an `actionAngleIsochrone` instance and by numerical integration using an `actionAngleSpherical` object. The radial and vertical actions for this orbit are 0.05 and 0.025, respectively. The radial action is conserved to 1 part in $10^{-8}$.

FIG. 16.— Error in the angle coordinates as a function of integration time for the same orbit and the same action–angle methods as in Figure 15. The error is computed as $\Delta\theta = \theta(t) - \theta(t=0) - \Omega\,t$, where the frequency $\Omega$ is calculated as the mean frequency over the orbit. The errors on the frequencies for both methods are $\approx 2 \times 10^{-9}$.

§ 2.2) is again recommended.

The relative-energy error during the orbit integration in `MWPotential2014` of the orbit displayed on the right in Figure 10 is shown in Figure 13. The energy error remains small for all of the integrators for thousands of orbits, with only `odeint` and `rk4_c` approaching relative-energy errors of $10^{-5}$ after 2000 periods. The energy error for the symplectic integrators does not grow with time as expected for phase–space-volume-conserving integrators. Because most of the phase–space of galaxies is at most a few hundred orbital times old at the present time, these energy errors are innocuous.

`galpy` also has the ability to integrate phase–space volumes $(\Delta\mathbf{x}, \Delta\mathbf{v})$ rather than just positions $(\mathbf{x}, \mathbf{v})$ through the `Orbit` method `integrate_dxdv`. The phase–space volume is integrated directly, that is, by integrating the equations of motion for the phase–space volume. This requires the second derivatives of the potential to be implemented for the potential in which the volume is being integrated. This functionality is currently only supported for two-dimensional orbits, because it is used for calculating the properties of two-dimensional DFs (see § 6.1). Extending this functionality to three-dimensional orbits would be straightforward and is planned for a future version of the code. As an example of this, Figure 14 shows how well the phase–space volume is conserved (Liouville's theorem) for the four non-symplectic integrators (symplectic integrators cannot be used for integrating $(\Delta\mathbf{x}, \Delta\mathbf{v})$ as the forces involved are not conser-vative). What is shown is the determinant of the Jacobian of the transformation between $(\Delta\mathbf{x}, \Delta\mathbf{v})(t=0)$ and $(\Delta\mathbf{x}, \Delta\mathbf{v})(t)$; this determinant should be one if phase–space volume is conserved by the integrator. The fourth- and sixth-order Runge–Kutta order methods perform best for the orbit displayed in Figure 14, with much faster-growing deviations for the `dopr54_c` and `odeint` solvers. The latter in particular performs poorly.

The various integrators contained in `galpy` are accessed in a uniform way (particularly the C solvers) and it is therefore easy to implement alternative methods. If these routines are cast in the same form as the currently implemented methods, adding them to the `galpy` framework is only a matter of adding and editing a few lines of code.

## 5. ACTION-ANGLE COORDINATES

### 5.1. Generalities

The calculation of action–angle coordinates is an integral part of `galpy`. They can be calculated for `Orbit` instances using a variety of approximations for different kinds of potentials and are used as part of some of the distribution functions described in § 6. Action–angle routines are available in the `galpy.actionAngle` module. Each different method is implemented as a class, e.g., `actionAngleSpherical` (see below) that has the following methods: (a) `__call__`, which computes the actions only, (b) `actionsFreqs` for calculation of the actions and the orbital frequencies, and (c) `actionsFreqsAngles`, which also computes the angles. The methods are arranged in this way, because the calculation of the frequencies typically involves the computation of the actions and, similarly, computing the angles typically requires the frequencies. Grouping the outputs as is done in the three methods therefore minimizes unnecessary duplica-

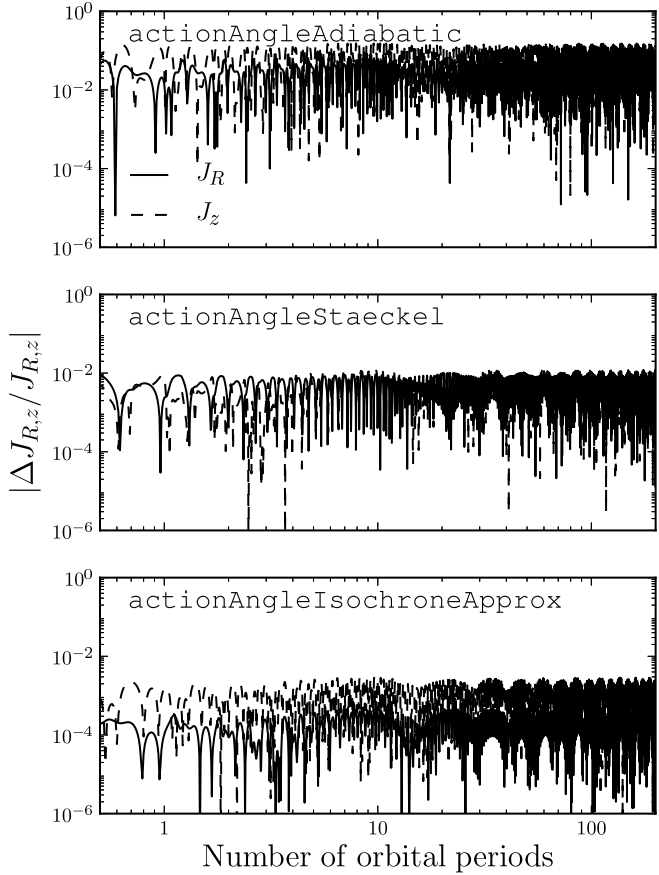

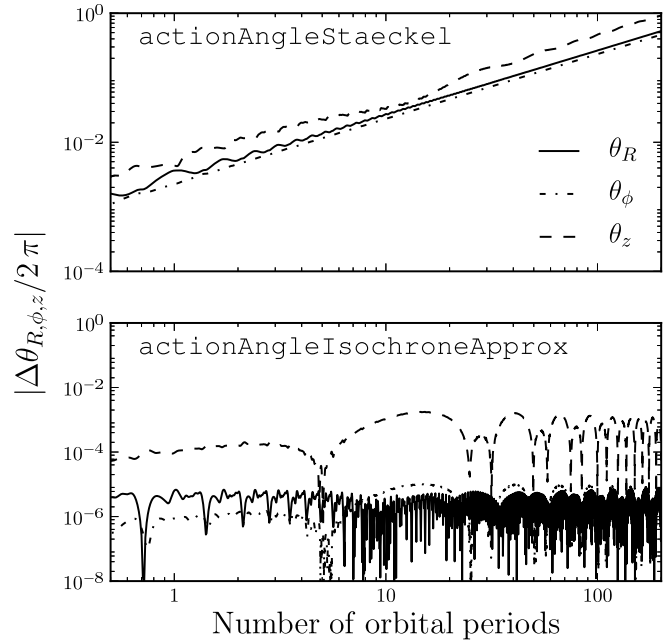

FIG. 17.— Variation of the radial and vertical actions as a function of integration time for axisymmetric potentials. The actions are computed at each point along the orbit displayed on the right in Figure 10. The actions are calculated using three different approximations: (a) the adiabatic approximation `actionAngleAdiabatic`, (b) the Stäckel approximation `actionAngleStaeckel`, and (c) a general orbit-integration-based method `actionAngleIsochroneApprox`. The radial and vertical actions for this orbit are 0.05 and 0.009, respectively. The Stäckel approximation conserves the radial action ≈ 5 times better than the adiabatic approximation and the vertical action ≈ 10 times better. The `actionAngleIsochroneApprox` with its default settings performs about 20 and 5 times better than the Stäckel approximation. All methods agree on the actions to within these errors.

FIG. 18.— Error in the angle coordinates with respect to the initial angle as a function of integration time for the same orbit and the same action–angle methods as in Figure 17 (except for `actionAngleAdiabatic` for which the frequencies and angles are currently not implemented). The error is computed as $\Delta\theta = \theta(t) - \theta(t=0) - \Omega\, t$, where the frequency $\Omega$ is calculated as the mean frequency over the orbit. The bottom panel covers a twice-as-large range on the $y$ axis. The errors on the radial, azimuthal, and vertical frequencies of the `actionAngleStaeckel` are $\approx 5 \times 10^{-5}$, $10^{-3}$, and $2 \times 10^{-3}$, respectively; for the `actionAngleIsochroneApprox` they are $\approx 4 \times 10^{-7}$, $10^{-6}$, and $7 \times 10^{-5}$, respectively. The smaller angle and frequency errors for `actionAngleIsochroneApprox` lead to angles that are stable to better than 1 in part $10^{-3}$ for hundreds of orbital periods.

### 5.2. Isochrone and spherical potentials

Action–angle coordinates and orbital frequencies for the isochrone potential, which can be calculated analytically, and for spherical potentials, which can be calculated using a few simple numerical integrals, are implemented in `galpy`. The isochrone routines are contained in the `actionAngleIsochrone` class, which is initialized using a specific isochrone potential. This can be done either by specifying the scale parameter of the isochrone potential, in which case the potential is assumed to be in natural coordinates (i.e., the circular velocity of the isochrone potential is 1 at $R = 1$). Alternatively, an instance of `IsochronePotential` can be passed.

Action–angle coordinates for spherical potentials are contained in the `actionAngleSpherical` class. Instances of this class are initialized by specifying a spherical potential. Note that whether or not the potential is actually spherical is not explicitly checked by the code, so care must be taken not to use it with axisymmetric potentials. The numerical integrals can be computed either by using `scipy`'s `integrate.quad` function or by using `scipy`'s `integrate.fixed_quad` routine; the latter uses Gaussian quadrature and is much faster.

The performance of the isochrone and spherical action–angle methods is demonstrated in Figures 15 and 16. Figure 15 shows the radial action computed using `actionAngleIsochrone` and `actionAngleSpherical` for a similar orbit as the one displayed on the right in Fig-tion in computations. Unless stated otherwise, all of the different types of action–angle calculations described below support all three of these methods.

The input to all three basic action–angle methods is the phase–space position $(\mathbf{x}, \mathbf{v})$. This can be specified in two different ways. The first is to pass an `Orbit` instance. The initial condition of this instance will be used as the phase–space position, unless a time is specified as well; in that case, the phase–space position at that time (if the orbit is integrated) will be used instead. Phase–space positions can also be specified directly in cylindrical Galactocentric coordinates. In this way, multiple phase–space positions can be passed at the same time, which is especially useful for those action–angle routines that are implemented in C, for which passing multiple points at once is more efficient.

`galpy` currently does not contain any methods for calculating positions and velocities for a given set of action–angle coordinates in a given potential.

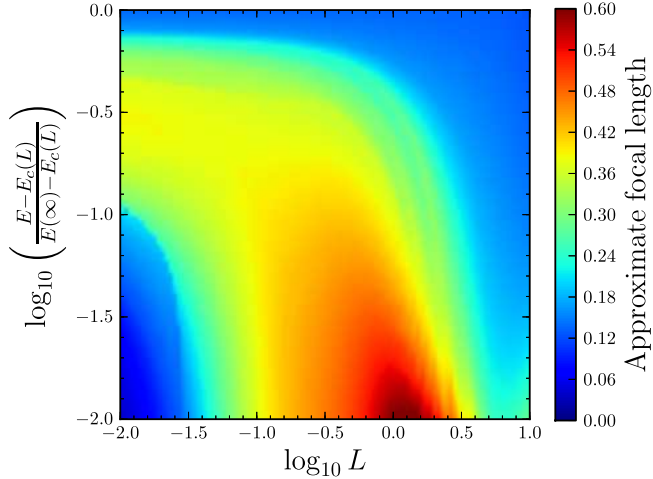

ure 10. The orbit is similar in that it has the same initial condition, but it is integrated in an isochrone potential with scale parameter 0.3; this isochrone potential is similar to `MWPotential2014` over the radial range covered by the orbit in Figure 10. The radial action is conserved to 1 part in $10^{-8}$ using both approximations. Figure 16 displays the error in the radial and vertical angles. What is shown is the difference between the angle at time $t$ and the angle predicted from the linear increase of initial angle at time $t = 0$ with the frequency calculated as the average frequency over the orbit. Thus, this error shows how consistent angles calculated at a later time are with the initial angle and frequency. The angle error itself is $\approx 10^{-8}$; the frequency error of $\approx 2 \times 10^{-9}$ leads to a linear rise in the angle error after more than a few orbital times, when frequency errors have had time to build up.

### 5.3. Action-angle coordinates for axisymmetric potentials

`galpy` also contains multiple approximations to the actions, angles, and frequencies for axisymmetric potentials. The first of these is the *adiabatic approximation* (e.g., Binney 2010). In this approximation, the motion of a star near the $z = 0$ symmetry plane of an axisymmetric potential is approximated as being decoupled oscillators in the radial and vertical part of the effective potential. This is a good approximation for stars on nearly-circular orbits that do not stray too far from the mid-plane of the potential; it is implemented as the `actionAngleAdiabatic` class, which is initialized using a potential instance and the specification of the $\gamma$ "fudge" factor of Binney & McMillan (2011).

The second approximation for axisymmetric potentials is the Stäckel approximation of Binney (2012). This method calculates action-angle coordinates by locally approximating the potential as a Stäckel potential specified by the focal length $\delta$ of a prolate spheroidal coordinate system. As shown by Binney (2012), by specifying

$\delta$, one can calculate approximate actions, frequencies, and angles without explicitly fitting a Stäckel potential to the axisymmetric potential in question. This allows action–angle coordinates to be calculated as efficiently as for the adiabatic approximation, using just a few one-dimensional numerical integrals. This Stäckel method is implemented as the `actionAngleStaeckel` class, which is instantiated by giving a potential instance and a focal length $\delta$. See § 3.1 of Bovy et al. (2013) for further information on the Stäckel approximation and how it is implemented in `galpy`.

Both the adiabatic and Stäckel approximations are implemented (partially) in `python` as well as in `C`. The actions can be calculated in `python` using `scipy`'s `integrate.quad` function or by using `scipy`'s `integrate.fixed_quad` routine; the latter is again much faster. The actions for both methods can also be computed in `C`, using Gaussian quadrature with 20 points. The frequencies and angles are currently not implemented for `actionAngleAdiabatic`, but they are available for `actionAngleStaeckel`. However, they are only implemented in `C`, and can therefore only be used with potentials that have `C` implementations. An interpolated potential that can be passed to `C` can be built using `interpRZPotential` (see § 3.3) if frequencies and angles are desired for axisymmetric potentials without `C` implementations.

The performance of the axisymmetric approximations are demonstrated in Figures 17 and 18 for the orbit shown on the right in Figure 10 in the `MWPotential2014` potential. For this orbit, `actionAngleAdiabatic` conserves the radial and vertical actions to a few percent, while the more precise `actionAngleStaeckel` display action fluctuations that are less than one percent. The consistency of the angles along the orbit with the initial angle plus the linear frequency×time increase is given in Figure 18. The angle error itself is $\approx 10^{-3}$; the inconsistency between the initial angle and the angle at time $t$ grows with time, but only becomes of order one after a few hundred periods. These errors are typical for disk orbits near the Sun in the Milky Way.

As described above, `actionAngleStaeckel` requires the user to specify the focal length of a prolate spheroidal coordinate system. An appropriate value for this focal length can be calculated using equation (9) in Sanders (2012), which gives the focal length of a true Stäckel potential in terms of the forces and second derivatives of the potential. If this is computed for a non-Stäckel potential, then an approximate value is found. A function `estimateDeltaStaeckel` included in `galpy.actionAngle` calculates this approximate focal length for a given position in an axisymmetric potential. If multiple positions are given, e.g., positions along an orbit, then the median of the individual focal lengths is returned. This function can be used to establish an approximate focal length for the Stäckel approximation for a given (orbit,potential) pair. In Figure 19, the approximate focal length thus computed for a wide range of orbits in `MWPotential2014` is shown as a function of energy and angular momentum. This figure can be used pick $\delta$ when using `actionAngleStaeckel` for `MWPotential2014`. Very close to and very far from the center, the potential is approximately spherical, because the bulge and halo are represented with spherical mod-els; therefore the focal length is close to zero. In regions where the disk dominates, $\delta \approx 0.3$ to $0.6$. It is clear that $\delta$ does not vary strongly between nearby orbits, such that a single $\delta$ often suffices for a wide range of orbits near a given position.

### 5.4. Grid-based action-angle coordinates for axisymmetric potentials

To speed up action calculations, both the adiabatic and the Stäckel methods can be used to tabulate the values of the radial and vertical actions on a grid in (approximate) integrals of the motion that can be interpolated for subsequent action evaluations. These grid-based methods are available as `actionAngleAdiabaticGrid` and `actionAngleStaeckelGrid`.

For `actionAngleAdiabaticGrid`, two separate grids for $J_z$ and $J_R$ are formed. The former is a grid in $(R, E_z)$, where $E_z$ is the vertical energy (see the caption of Figure 11 for a precise definition), because $J_z$ only depends on $R$ and $E_z$ in the adiabatic approximation. The grid in $R$ is regular over $0.01 \leq R \leq R_{\max}$. At each $R_i$, we then calculate $E_{z,\max}(R_i) \equiv E_z(z_{\max}; R_i)$, evaluate $J_z$ on a regular grid between zero and $E_z(z_{\max}; R)$, and divide by $J_{z,\max}(R_i) \equiv J_z(R_i, E_{z,\max}(R_i))$. We then employ third-degree bivariate spline interpolation of this normalized $J_z$ and third-degree one-dimensional spline interpolation to interpolate the functions $E_{z,\max}(R)$ and $J_{z,\max}(R)$ as a function of $R$. To evaluate $J_z$ for a given orbit (essentially $[\tilde{R}, \tilde{z}, \tilde{v}_z]$) using this grid, we compute $\tilde{E}_z(\tilde{R}, \tilde{z}, \tilde{v}_z)$, divide by the appropriate $E_{z,\max}(\tilde{R})$, and evaluate $J_z(\tilde{R}, \tilde{E}_z/E_{z,\max}(\tilde{R}))/J_{z,\max}(\tilde{R})$ using the bivariate spline. By multiplying this by $J_{z,\max}(\tilde{R})$, $J_z(\tilde{R}, \tilde{z}, \tilde{v}_z)$ is finally obtained. The keyword parameters `Rmax`, `zmax`, `nR`, and `nEz` set the parameters $R_{\max}$ and $z_{\max}$ and the size of the grid.

$J_R$ only depends on $L_z$ and $E_R$—the radial energy, see the caption of Figure 11—in the adiabatic method. We build a regular grid in $L_z$ over $0.01 \leq L_z \leq v_c(R_{\max})R_{\max}$, using the same $R_{\max}$ as above. We also calculate the radius $R_L$ of a circular orbit with angular momentum $L_{z,i}$ at each grid point $L_{z,i}$. Similar to the $J_z$ grid above, at each $L_{z,i}$ we then compute $E_{R,\min,i}(L_z) \equiv E_R(v_R = 0, v_T = L_{z,i}/R_{L,i}, R = R_{L,i})$ and $E_{R,\max,i}(L_z) \equiv E_R(v_R = 0, v_T = L_{z,i}/R_{\max}, R = R_{\max})$ and form a regular grid in $E_R$ at each $L_{z,i}$ between $E_{R,\min,i}(L_{z,i})$ and $E_{R,\max,i}(L_{z,i})$. We again use third-degree bivariate spline interpolation to interpolate the normalized $J_R$ and third-degree one-dimensional spline interpolation to interpolate $E_{R,\min}(L_z)$, $E_{R,\max}(L_z)$, and $J_{R,\max}(L_{z,i})$ as functions of $L_z$. To evaluate $J_R$ for a new orbit (essentially $[\tilde{R}, \tilde{v}_R, \tilde{v}_T]$), we calculate $\tilde{L}_z$—as $\tilde{R}\tilde{v}_T + \gamma J_z$ using the fudge $\gamma$—and $\tilde{E}_R$ and then follow a procedure similar to that above for $J_z$. The size of the grid is controlled by the parameters `nEr` and `nLz`.

To interpolate the Stäckel method, we roughly follow the procedure described in § 2.2 of Binney (2012) with some tweaks for efficiency. Because it is beyond the scope of this paper to fully describe the Stäckel method, I will employ the terminology of Binney (2012); the reader should refer to that paper for full details. We build a regular three-dimensional grid, starting with $L_z$ over $0.01 \leq L_z \leq v_c(R_{\max})R_{\max}$, where $R_{\max}$ is again set as an input parameter `Rmax`. We also calculate the radius $R_L$ of a circular orbit with angular momentum $L_z$ at each grid point $L_{z,i}$. At each $L_{z,i}$, we then compute the energy of a circular orbit $E_c(L_{z,i})$ and the energy $E_{\max}(L_{z,i})$ of an orbit with $z = v_z = v_R = 0$ and $R = 25$ similar to how this is done for the adiabatic grid above (this arbitrary $R = 25$ constant assumes that we are working in galpy's natural coordinates; see § 2.2) and we build a logarithmic grid between these extreme energies to accurately capture orbits ranging from very-close-to-circular to radial. The Stäckel algorithm uses a reference value $u_0$ for the equivalent of the radial coordinate in the prolate spheroidal coordinate system. While the value of $u_0$ does not matter for direct evaluations of the Stäckel algorithm, it does matter for the gridded evaluation below. This is essentially because we build the grid in approximate integrals of the motion and for a good value of $u_0$, these integrals are better. We determine $u_0$ by minimizing the expression in equation (20) of Binney (2012) on the grid in $L_z$ and $E$ described above and interpolate its natural logarithm using third-degree bivariate spline interpolation.

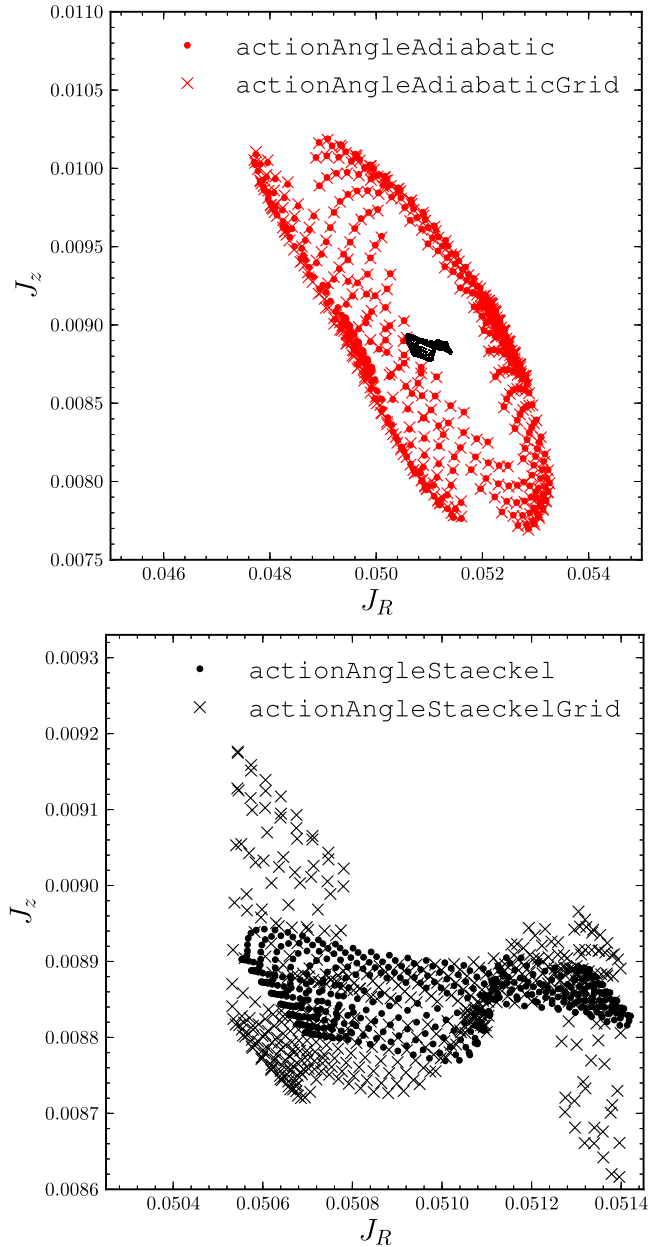

FIG. 20.— Vertical vs. radial action over five orbital periods along the orbit displayed on the right in Figure 10, calculated with the axisymmetric action–angle methods. The top panels shows this for actions calculated using the adiabatic approximation as well as for its grid-based-interpolation version. The bottom panel shows the actions computed with the Stäckel method and its grid-based-interpolation version. The direct Stäckel actions are also shown as black points in the top panel to emphasize that the actions are much better conserved using the Stäckel method. Because the adiabatic actions only depend on the approximate integrals of the motion that are used to tabulate the actions for interpolation, the interpolated values agree exactly with the direct calculation (to within the interpolation errors; the crosses and dots almost exactly overlap). As the tabulation of the Stäckel method uses approximate integrals that are *not* depended on in the direct calculation, the grid-based values deviate from the direct values. In particular, the grid-based $J_z$ varies by about a factor of three more than the direct $J_z$.

We then determine the speed $w$ that a phase-space point at $(L_z, E)$ and $(u, v) = (u_0, \pi/2)$ has (where $v$ is the 'vertical' coordinate in the spheroidal coordinate system) and calculate the radial and vertical action on a three-dimensional grid in $(L_z, E, \psi)$, where $L_z$ and $E$ are the same grid as above and $\psi$ is a regular grid between 0 and $\pi/2$. These actions are calculated for a phase-space point $(R, vR, vT, v_T, z, v_z) = (\Delta \sinh u_0, w \cos \psi, L_z/\Delta/\sinh u_0, 0, w \sin \psi)$. Similar to the adiabatic grid above, we divide each $J_R$ and $J_z$ two-dimensional sub-grid at $L_{z,i}$ by its maximum and we then use three-dimensional third-degree spline interpolation to interpolate these normalized actions and one-dimensional spline interpolation to interpolate the maxima as a function of $L_z$. This gridding approach is slightly different from that of Binney (2012), who uses a grid in $(L_z, E, E_r)$, where $E_r$ is a radial energy given by equation (18) of Binney (2012).

To evaluate the actions for a new position $(\tilde{R}, \tilde{v}_R, \tilde{v}_T, \tilde{z}, \tilde{v}_z)$, we first calculate $\tilde{L}_z$ and $\tilde{E}$ and use the interpolation grid for $u_0$ to determine $\tilde{u}_0$. We then need to determine the appropriate value of $\tilde{\psi}$ to use the interpolation grid in $(L_z, E, \psi)$. We determine two separate values for $\tilde{\psi}$ for the purpose of evaluating $J_R$ and $J_z$. We compute $\tilde{\psi}_R$ by calculating $\tilde{E}_r$ and the velocity $\tilde{w}$ of the phase-space point at $(\tilde{L}_z, \tilde{E})$ and $(u, v) = (\tilde{u}_0, \pi/2)$ and determining $\tilde{\psi}_R$ from

$$\cos^2 \tilde{\psi}_R = \frac{2\tilde{E}_r}{\tilde{w}^2 \cosh^2 \tilde{u}_0}. \tag{1}$$

$\tilde{J}_R$ is then found using the interpolated $J_R$ grid evaluated at $(\tilde{L}_z, \tilde{E}, \tilde{\psi}_R)$.

For $\tilde{J}_z$ we proceed similarly, but we use a vertical energy $E_z$, similar to the radial energy $E_r$, defined as

$$E_z = \frac{p_v^2}{2\Delta^2} + \frac{L_z^2}{2\Delta^2}\left(\frac{1}{\sin^2 v - 1}\right) - E(\sin^2 v - 1) - \delta V(v), \tag{2}$$

using the same notation as Binney (2012). We compute

this $\tilde{E}_z$, find $\tilde{\psi}_z$ as

$$\sin^2 \tilde{\psi}_z = \frac{2\tilde{E}_z}{\tilde{w}^2 \cosh^2 \tilde{u}_0}, \tag{3}$$

and use $(\tilde{L}_z, \tilde{E}, \tilde{\psi}_z)$ to evaluate $J_z$.

The actions along the orbit displayed on the right in Figure 10 computed with the direct adiabatic and Stäckel methods and with their grid-based-interpolation versions are demonstrated in Figure 20.

If the potential were an exact Stäckel potential, $E_r$ and $E_z$ would both be exact integrals of the motion and $\psi_R$ would be exactly equal to $\psi_z$. Typical galactic potentials are not exact Stäckel potentials, so this will only hold approximately. The reason that $\psi_z$ is a better $\psi$ to use in the interpolation of $J_z$ is because $E_z$ is closely related to the constant of the motion $I_3 + V(\pi/2)$ used in the direct calculation of $J_z$ using the Stäckel method. $E_r$ has the same relation to the constant $I_3 + U(u_0)$, used in the direct calculation of $J_R$. For non-Stäckel potentials, these two constants are not interchangeable, such that using $\psi_z$ in the interpolation gives a more accurate value of $J_z$. As a specific example, if we had used $\psi_R$ instead of $\psi_z$ to compute $J_z$ in the bottom panel of Figure 20, the interpolated $J_z$ would follow a straight line from the top-left to the bottom-right, because errors in $J_R$ would directly propagate into errors in $J_z$, leading to a complete degeneracy.

### 5.5. Action-angle coordinates for general static potentials

A general method for calculating action–angle coordinates for a static potential is contained in the `actionAngleIsochroneApprox` class. This method works by calculating action–angle coordinates in an auxiliary isochrone potential at each point along the orbit. Actions, frequencies, and angles are then obtained by fitting—implicitly for the actions—a generating function between the (incorrect) action–angle coordinates of the isochrone potential and the (correct) action–angle coordinate of the potential of interest (see the appendix of Bovy 2014 for full details). The `actionAngleIsochroneApprox` methods are built entirely on the framework provided by `galpy.potential`, `galpy.orbit`, and `galpy.actionAngle.actionAngleIsochrone`.

The performance of `actionAngleIsochroneApprox` is demonstrated in Figures 17 and 18 for the orbit in `MWPotential2014` on the right in Figure 10. The actions are conserved to better than 1 part in $10^{-3}$. The error in the angle with respect to the initial angle plus linear increase remains small for hundreds of periods. This is essentially due to the fact that the frequency is fit simultaneously with the initial angle to the angles in the auxiliary isochrone potential over many orbital periods. The stability of the `actionAngleIsochroneApprox` algorithm makes it ideal for investigating the behavior of the angles over many orbital periods, such as for tidal streams (Bovy 2014).

The `actionAngleIsochroneApprox` method for calculating frequencies and angles is not implemented for non-axisymmetric potentials in the current version of `galpy`.

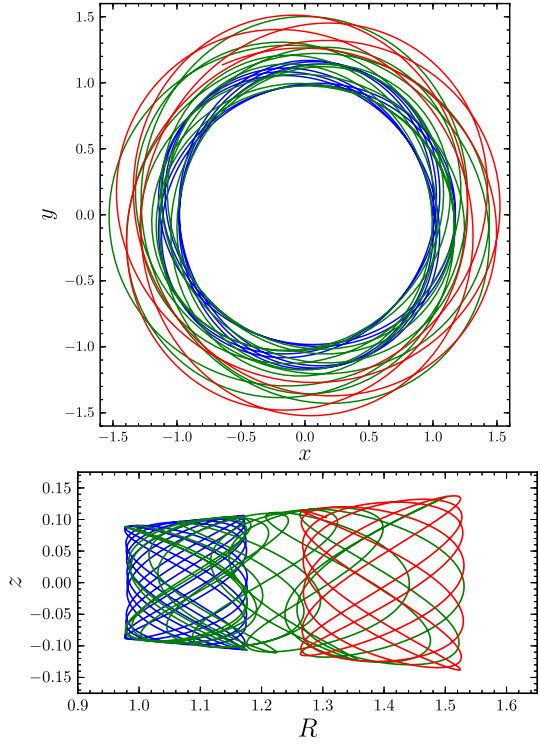

```
1  from galpy.potential import IsochronePotential
2  from galpy.orbit import Orbit
3  from galpy.actionAngle import actionAngleIsochrone
4  # Initialize two different IsochronePotentials
5  ip1= IsochronePotential(normalize=1.,b=1.)
6  ip2= IsochronePotential(normalize=0.5,b=1.)
7  # Use TimeInterpPotential to interpolate smoothly
8  tip= TimeInterpPotential(ip1,ip2,dt=100.,tform=50.)
9  # Integrate: 1) Orbit in the first isochrone potential
10 o1= Orbit([1.,0.1,1.1,0.0,0.1,0.])
11 ts= numpy.linspace(0.,50.,1001)
12 o1.integrate(ts,tip)
13 o1.plot(d1='x',d2='y',xrange=[-1.6,1.6],yrange=[-1.6,1.6],
14     color='b')
15 # 2) Orbit in the transition
16 o2= o1(ts[-1]) # Last time step => initial time step
17 ts2= numpy.linspace(50.,150.,1001)
18 o2.integrate(ts2,tip)
19 o2.plot(d1='x',d2='y',overplot=True,color='g')
20 # 3) Orbit in the second isochrone potential
21 o3= o2(ts2[-1])
22 ts3= numpy.linspace(150.,200.,1001)
23 o3.integrate(ts3,tip)
24 o3.plot(d1='x',d2='y',overplot=True,color='r')
25 # Now we calculate energy, maximum height, and mean radius
26 print o1.E(pot=ip1), (o1.rperi()+o1.rap())/2, o1.zmax()
27 -2.79921356237 1.07854158141 0.106331362938
28 print o3.E(pot=ip2), (o3.rperi()+o3.rap())/2, o3.zmax()
29 -1.19677002624 1.39962036137 0.138364269321
30 # The orbit has clearly moved to larger radii,
31 # the actions are however conserved from beginning to end
32 aAI1= actionAngleIsochrone(ip=ip1); print aAI1(o1)
33 (array([ 0.00773779]), array([ 1.1]), array([ 0.0045361]))
34 aAI2= actionAngleIsochrone(ip=ip2); print aAI2(o3)
35 (array([ 0.00773812]), array([ 1.1]), array([ 0.0045361]))
```

FIG. 21.— Adiabatic invariance of the actions. This figure uses galpy's action–angle routines to demonstrate that the actions are conserved when adiabatically deforming one isochrone potential into another, using TimeInterpPotential defined in Figure 4. In the figures, the blue lines display the orbit in the first potential, the green lines the orbit while the potential is being changed, and the red lines the orbit in the second potential. The top figure shows the orbit projected onto the mid-plane and the bottom figure shows the orbit in the meridional plane.

### 5.6. Example: Adiabatic invariance of the actions

To illustrate the use of the action–angle routines in practice, Figure 21 demonstrates that the actions are conserved when adiabatically changing the potential. The code is given in the bottom part of the figure; it makes use of TimeInterpPotential of Figure 4 to perform the smooth change between two isochrone potentials. The energy, mean radius, and maximum height reached above the plane clearly change, but the actions are conserved.

## 6. DISTRIBUTION FUNCTIONS FOR DISKS

Three classes of DFs for axisymmetric disks are currently included in galpy. Two of these are purely two-dimensional models: the classical Shu distribution (Shu 1969) and the 'new' disk DF of Dehnen (1999) (which we refer to as a Dehnen DF). These are described in § 6.1. The third family of disk DFs are the quasi-isothermal DFs first proposed by Binney (2010) and later refined by Binney & McMillan (2011); this is a family of fully three-dimensional DFs. The Shu and Dehnen DFs use the energy $E$ and $z$-component of the angular momentum $L_z$ as the arguments of the DF, while the quasi-isothermal DF uses the orbital actions.

### 6.1. Two-dimensional distribution functions

galpy contains both the Shu and Dehnen DFs described in Dehnen (1999) under galpy.df; both are implemented as subclasses of a general galpy.df.diskdf. These are two-dimensional, steady-state, axisymmetric DFs that use $E$ and $L_z$ as the sole arguments of the DF. They are both warmed up versions of a kinematically-cold disk DF consisting of circular orbits only with some surface-density profile $\Sigma(R)$; they differ in how this cold DF is warmed-up, i.e., generalized to include non-circular orbits.

The classical Shu DF is given by

$$f_{\text{Shu}}(E, L_z) = \frac{\Omega(R_L)\Sigma(R_L)}{\pi\kappa(R_L)\sigma_R^2(R_L)} \exp\left[\frac{E_c(L_z) - E}{\sigma_R^2(R_L)}\right],$$
(4)

where $\Omega$ is the rotational frequency, $\kappa$ is the epicycle frequency, $\Sigma(\cdot)$ and $\sigma_R(\cdot)$ are scale surface-density and radial-velocity-dispersion profiles; all of these functions are evaluated at $R_L$, the radius of a circular orbit with angular momentum equal to $L_z$. The quantity $E_c(L)$ is the energy of this circular orbit. The Shu DF is available as galpy.df.shudf.

The Dehnen DF has the following form

$$f_{\text{Dehnen}}(E, L_z) =$$
$$\frac{\Omega(R_E)\Sigma(R_E)}{\pi\kappa(R_E)\sigma_R^2(R_E)} \exp\left[\frac{\Omega(R_E)\left(L - L_c(E)\right)}{\sigma_R^2(R_E)}\right].$$
(5)

Here, $R_E$ is the radius of a circular orbit with energy $E$ and all of the functions appearing in this DF, which are the same as those appearing in the Shu DF, are evaluated at $R_E$ rather than at $R_L$. The Dehnen DF is available as galpy.df.dehnendf.

We refer the reader to Dehnen (1999) for a detailed discussion of the advantages and disadvantages of these two DFs and of their properties. As discussed by Dehnen (1999), the surface-density $\Sigma_{\text{DF}}$ and radial-velocity-dispersion profiles $\sigma_{R,\mathrm{DF}}$ corresponding to the Shu and Dehnen DFs differ from the scale profiles $\Sigma(R)$ and $\sigma_R(R)$ that appear in the definitions of these DFs. In § 3.2 of Dehnen (1999), a procedure is given to correct the scale profiles to obtain a desired set of $(\Sigma_{\mathrm{out}}(R), \sigma_{R,\mathrm{out}}(R)^2)$—typically these are exponential profiles characterized by a scale length and normalization at $R_0$—using an iterative procedure that starts with $\Sigma(R) = \Sigma_{\mathrm{out}}$ and $\sigma_R = \sigma_{R,\mathrm{out}}$ and applies multiplicative corrections to these based on the difference $\Sigma_{\mathrm{DF}}/\Sigma_{\mathrm{out}}$ and $\sigma_{R,\mathrm{DF}}/\sigma_{R,\mathrm{out}}$. This procedure is implemented in **galpy** in a **DFcorrection** class. The user can specify the number of iterations to use and correction factors are calculated and stored automatically for re-use. Corrections are stored, because the iterative procedure is quite slow[3]. A large number of correction factors stored in the **galpy** format are available for download online for a few profiles $(\Sigma_{\mathrm{out}}(R), \sigma_{R,\mathrm{out}}(R)^2)$. Instructions on how to download and install these are given in the online documentation.

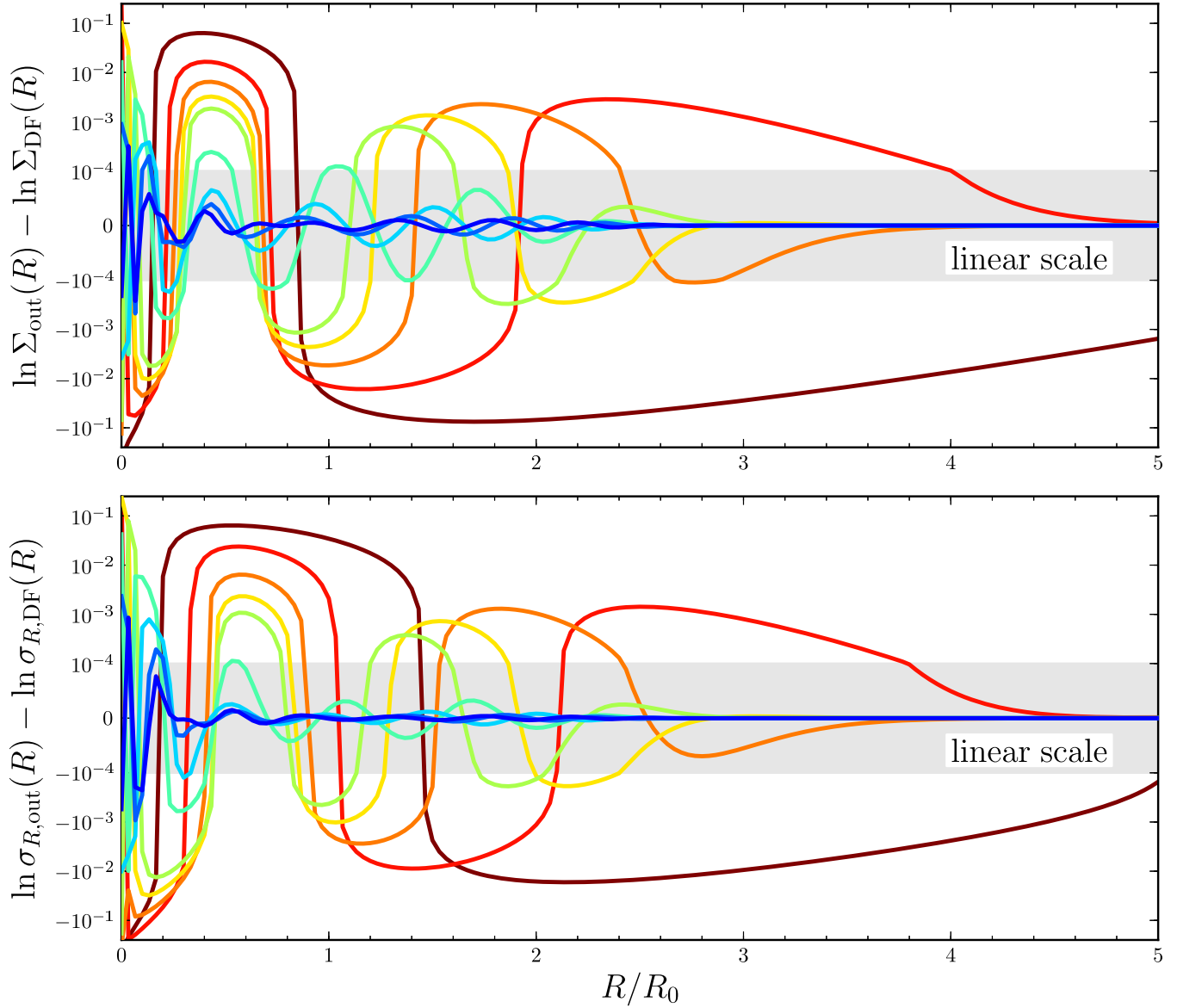

FIG. 22.— Logarithmic difference between the desired surface-density profile $\Sigma_{\mathrm{out}}(R)$ and the surface-density $\Sigma_{\mathrm{DF}}(R)$ obtained by integrating the DF over velocity for a **dehnendf** in a logarithmic potential with $\Sigma_{\mathrm{out}}(R) \propto \exp(-R/(R_0/3))$ and $\sigma_{R,\mathrm{out}}(R) = 0.2 \exp(-(R - R_0)/R_0)$ (*top panel*). The difference is shown after 1, 2, 3, 4, 5, 10, 15, 20, and 25 iterations of the correction-procedure of Dehnen (1999) (dark red through dark blue curves). The bottom panel shows the logarithmic difference between $\sigma_{R,\mathrm{out}}(R)$ and $\sigma_{R,\mathrm{DF}}(R)$. The $y$ axis is logarithmic, except for the gray band. This figure shows that the correction procedure is highly effective for generating a DF that has a desired set of $(\Sigma_{\mathrm{out}}(R), \sigma_{R,\mathrm{out}}(R)^2)$: logarithmic differences $\lesssim 10^{-2}$ and $\lesssim 10^{-5}$ are obtained after 3 and 15 iterations, respectively.

Figure 22 demonstrates how the procedure of correcting the $\Sigma(R)$ and $\sigma_R(R)$ profiles in the DF leads to the desired set of $(\Sigma_{\mathrm{out}}(R), \sigma_{R,\mathrm{out}}(R)^2)$ for $\Sigma_{\mathrm{out}}(R) \propto \exp[-R/h_R]$ and $\sigma_{R,\mathrm{out}}(R) = \sigma_R(R_0) \exp[-(R - R_0)/h_\sigma]$, with $h_R = R_0/3$, $h_\sigma = R_0$,

---

[3] Note that this is partially because **galpy** calculates the surface-density and radial-velocity dispersion at $R$ by direct two-dimensional integration of the DF over the velocity. However, in the case of the Shu DF these integrals could be re-written as one-dimensional integrals (see Sharma & Bland-Hawthorn 2013) and thus sped up.

```
1  from galpy.df import dehnendf
2  # Init. dehnendf w/ flat rot., hr=1/3, hs=1, and sr1()=0.2
3  df= dehnendf(beta=0.,profileParams=(1./3.,1.0,0.2))
4  # Same, w/ correction factors to scale profiles
5  dfc= dehnendf(beta=0.,profileParams=(1./3.,1.0,0.2),
6                correct=True,niter=20)
7  # Log. diff. between scale and DF surf. dens.
8  numpy.log(df.surfacemass(0.5)/df.targetSurfacemass(0.5))
9  -0.056954077791649592
10 # Same for corrected DF
11 numpy.log(dfc.surfacemass(0.5)/dfc.targetSurfacemass(0.5))
12 -4.1640377205802041e-06
13 # Log. diff between scale and DF sr
14 numpy.log(df.sigmaR2(0.5)/df.targetSigma2(0.5))
15 -0.1278608300136327
16 # Same for corrected DF
17 numpy.log(dfc.sigmaR2(0.5)/dfc.targetSigma2(0.5))
18 -6.8065001252214986e-06
19 # Evaluate DF w/ R,vR,vT
20 df(numpy.array([0.9,0.1,0.8]))
21 array(0.1740247246180417)
22 # Evaluate corrected DF w/ Orbit instance
23 from galpy.orbit import Orbit
24 dfc(Orbit([0.9,0.1,0.8]))
25 array(0.16834863725552207)
26 # Calculate the mean velocities
27 df.meanvR(0.9),  df.meanvT(0.9)
28 (0.0, 0.91144428051168291)
29 # Calculate the velocity dispersions
30 numpy.sqrt(dfc.sigmaR2(0.9)), numpy.sqrt(dfc.sigmaT2(0.9))
31 (0.22103383792719539, 0.17613725303902811)
32 # Calculate the skew of the velocity distribution
33 df.skewvR(0.9),  df.skewvT(0.9)
34 (0.0, -0.4733163836602586)
35 # Calculate the kurtosis of the velocity distribution
36 df.kurtosisvR(0.9),  df.kurtosisvT(0.9)
37 (-0.13561300880237059, 0.12612702099300721)
38 # Calculate a higher-order moment of the velocity DF
39 df.vmomentsurfacemass(1.,6.,2.)/df.surfacemass(1.)
40 0.0004895349220555950
41 # Calculate the Oort functions
42 dfc.oortA(1.), dfc.oortB(1.), dfc.oortC(1.), dfc.oortK(1.)
43 (0.40958989076012197, -0.49396172114486514, 0.0, 0.0)
44 # Sample Orbits from the DF, returns list of Orbits
45 os= dfc.sample(n=100,returnOrbit=True,nphi=1)
46 # check that these have the right mean radius = 2hr=2/3
47 rs= numpy.array([o.R() for o in os])
48 assert numpy.fabs(numpy.mean(rs)-2./3.) < 0.1
49 # Sample vR and vT at given R, check their mean
50 vrvt= dfc.sampleVRVT(0.7,n=500,target=True); vt= vrvt[:,1]
51 assert numpy.fabs(numpy.mean(vrvt[:,0])) < 0.05
52 assert numpy.fabs(numpy.mean(vt)-dfc.meanvT(0.7)) < 0.01
53 # Sample Orbits along a given line-of-sight
54 os= dfc.sampleLOS(45.,n=1000)
```

FIG. 23.— Methods of `diskdf` instances (`shudf` or `dehnendf`), illustrated using a Dehnen DF with a flat rotation curve ($\beta = 0$ in $v_c(R) = v_c(R_0) (R/R_0)^\beta$) with $\Sigma(R) \propto \exp[-R/(R_0/3)]$ and $\sigma_R(R) = 0.2 \exp[-(R - R_0)/R_0]$. The `df` object does not apply corrections to these profiles, while the `dfc` object does, with 20 iterations of the correction procedure (see text). The results on the moments of the uncorrected DF can be directly compared to Figure 4 of Dehnen (1999).

and $\sigma_R(R_0) = 0.2$ (in *natural coordinates*). This figure shows that the correction factors are highly effective in making the differences $\Sigma_{\mathrm{DF}}/\Sigma_{\mathrm{out}}$ and $\sigma_{R,\mathrm{DF}}/\sigma_{R,\mathrm{out}}$ extremely small.

The initialization of a `shudf` or a `dehnendf` instance requires one to specify the potential (through the `beta=` keyword that sets the logarithmic slope of the rotation curve, see below), the surface-density and radial-velocity-dispersion profiles $\Sigma(R)$ and $\sigma_R(R)$, and whether or not to apply corrections. When applying corrections, some keywords related to the iterative calculation of the correction factors can also be specified.

Many properties of the Shu and Dehnen DFs can be calculated by `galpy`. This includes the surface-density, mean velocities, velocity dispersions, as well as higher-order moments of the velocity DF. We can also calculate the Oort functions $A(R)$, $B(R)$, $C(R)$, and $K(R)$ (see, e.g., Kuijken & Tremaine 1994 for a definition of these in terms of mean velocities and their derivatives; this definition can be applied to kinematically-warm populations). These are calculated by integration over the DF and radial derivatives of the DF. As an axisymmetric DF, $C(R)$ and $K(R)$ should be exactly zero everywhere, but they are calculated by explicit integration as a check on the DF implementation. We can further sample from the DF in a variety of ways: sampling (a) $(E, L_z)$ or full $(R, v_R, v_T, \phi)$ phase-space coordinates with or without a restriction in radius (following the procedure given in § 5 of Dehnen 1999), (b) distances along a given line-of-sight, (c) $(v_R, v_T)$ velocities at a given position, or (d) full $(R, v_R, v_T, \phi)$ phase-space coordinates along a given line-of-sight. Most of the methods available for `dehnendf` and `shudf` instances are demonstrated in Figure 23.

A limitation of the current `diskdf` implementation is that it only supports power-law or logarithmic potentials, i.e., power-law rotation curves (including a flat rotation curve). While it would be relatively straightforward to generalize the code to use any `Potential` instance, this has not been done so far[4]. As the main purpose of the Shu and Dehnen DFs is to provide simple models for the kinematics of a disk galaxy, this is a not a serious limitation.

### 6.2. *Example: Oort functions for different tracer populations of stars*

As an example of the functionality in the `galpy.df.diskdf` module, I compute the Oort functions $A(R)$ and $B(R)$ at the solar radius[5] $R_0$ for populations with different radial-velocity dispersions and investigate how they depend on the asymmetric drift, i.e., the offset between the mean rotational velocity of a kinematically-warm population and the circular velocity (Binney & Tremaine 2008). This can be achieved in `galpy` by running

```
from galpy.df import dehnendf
df= dehnendf(beta=0.,correct=True,niter=20,
             profileParams=(1./3.,1.,0.1))
va= 1.-df.meanvT(1.) # asymmetric drift
A= df.oortA(1.)
B= df.oortB(1.)
```

for a fiducial model of a `dehnendf` with an exponential surface-density profile with scale length $h_R = R_0/3$, an exponential radial-velocity-dispersion profile with scale $h_\sigma = R_0$, and a flat rotation curve (`beta=0` for $\beta$ in $v_c(R) = v_c(R_0) (R/R_0)^\beta$). As usual, we work in natural

---

[4] Basically, the only functionality to be implemented is the calculation of $R_E$ for any potential (the calculation of $R_L$ for any potential is already a `Potential` method). The `diskdf` code would also have to be edited to make efficient use of $R_L$ and $R_E$. The quasi-isothermal DF implementation described below also depends on $R_L$ and that code does allow any `Potential` instance or list thereof to be used.

[5] Because there is no absolute scale in any of the ingredients of this calculation, the ratios $R_0/h_R$ and $R_0/h_\sigma$ are all that matters and the results below can be rescaled to other radii.

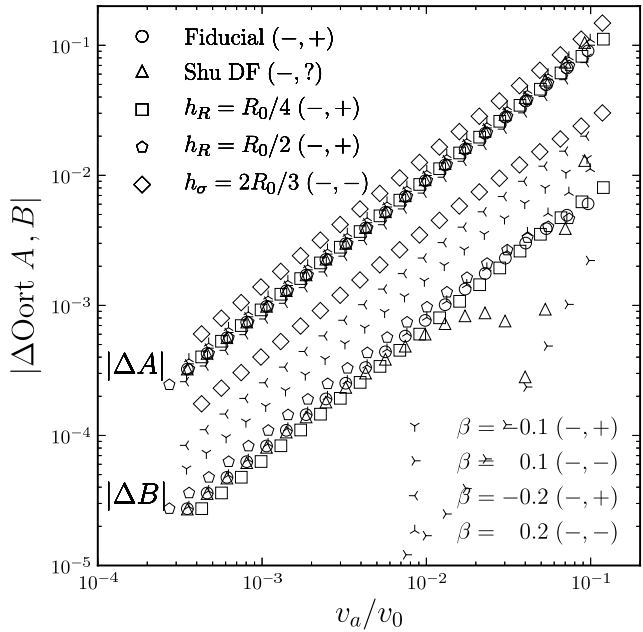

| | |
|---|---|
| ○ | Fiducial $(-,+)$ |
| △ | Shu DF $(-,?)$ |
| □ | $h_R = R_0/4$ $(-,+)$ |
| ○ | $h_R = R_0/2$ $(-,+)$ |
| ◇ | $h_\sigma = 2R_0/3$ $(-,-)$ |

$|\Delta A|$

$|\Delta B|$

$\vdash$  $\beta = 0.1$ $(-,+)$
$\dashv$  $\beta = 0.1$ $(-,-)$
$\prec$  $\beta = -0.2$ $(-,+)$
$\succ$  $\beta = 0.2$ $(-,-)$

FIG. 24.— Oort functions $A$ and $B$ for a warm axisymmetric disk. The Oort functions $A(R)$ and $B(R)$ as well as the asymmetric drift $(v_a(R) = v_c(R) - \bar{v}_T(R))$; $v_0 = v_c(R_0) = 1)$ are computed at $R_0$ for warm `diskdfs` as a function of the radial-velocity-dispersion at $R_0$. The difference between the Oort functions and the Oort constants, that is, the Oort functions for stars or gas on circular orbits, is displayed as a function of the asymmetric drift. The fiducial model is a `dehnendf` with an exponential surface-density profile with scale length $h_R = R_0/3$, an exponential radial-velocity-dispersion profile with scale length $h_\sigma = R_0$, and a flat rotation curve $(\beta = 0)$. Different symbols indicate variations on this fiducial model. In particular, the triangle uses the family of `shudf` DFs, and the rotation-curve parameter $\beta$ in $v_c(R) = v_c(R_0)\,(R/R_0)^\beta$ is varied over a wide range. The pair of '+' or '-' for each symbol indicates whether the difference whose absolute value is plotted is positive or negative for $A$ and $B$, respectively; '?' for the Shu DF indicates that the sign of $\Delta B$ can be both positive and negative: all but the final three points are positive. All of the DF variations except for the one with the shorter $h_\sigma$ display the same close-to-linear trend with $v_a$ for $\Delta A$ (this is difficult to see as many of the symbols overlap). The difference $\Delta B$ is also close to linear, but it depends more strongly on the form and parameters of the DF (all of the symbols including and below the lower sequence of diamonds are for $|\Delta B|$). For a population on circular orbits $A(v_a = 0) = (1-\beta)/2$ and $B(v_a = 0) = -(1+\beta)/2$ in these natural units. Even though $\Delta B$ strongly depends on the properties of the DF, it only differs by $\lesssim 5\%$ from the Oort constant $B(v_a = 0)$, even for populations with large asymmetric drift.

units in which $R_0$ and $v_0 = v_c(R_0)$ are both equal to one. This specific example calculates $A$ and $B$ for a population with a radial velocity dispersion at the Sun of $\sigma_R(R_0) = 0.1\,v_0$ and by varying this parameter we can investigate $A[\sigma_R(R_0)]$.

I compare $A[\sigma_R(R_0)]$ and $B[\sigma_R(R_0)]$ with the Oort constants, that is, $A[\sigma_R(R_0) = 0]$ and $B[\sigma_R(R_0) = 0]$, as a function of the asymmetric drift of the population. For the power-law rotation curve used here $A[\sigma_R(R_0) = 0] = (1-\beta)/2$ and $B[\sigma_R(R_0) = 0] = -(1+\beta)/2$. This comparison is presented in Figure 24 for DFs with different surface-density and velocity-dispersion profiles and with different forms (the `shudf`).

It is clear from Figure 24 that both $A$ and $B$ depend linearly on the asymmetric drift and that $A$ is much more affected by the kinematic temperature of the population than $B$. The effect of varying the DF form or parameters is much less pronounced on $A$ than it is on $B$, with the only parameter making a large difference for $A$ being $h_\sigma$. The behavior of $\Delta B$ for the Shu DF is more erratic than for the Dehnen DF, but $\Delta B$ remains small.

We can understand all of this behavior by calculating the dependence of $A$ and $B$ on the asymmetric drift using the radial, cylindrical Jeans equation. This equation for exponential surface-density and velocity-dispersion profiles can be written as

$$v_c^2 - \bar{v}_T^2 = \sigma_R^2\,\left(\gamma^2 + R\left[\frac{1}{h_R} + \frac{2}{h_\sigma}\right] - 1\right),\quad (6)$$

where $\gamma^2 = \sigma_T^2/\sigma_R^2$. Taking the derivative of this, we find using (a) that $\gamma^2 - 1$ is typically much smaller than the other term in the parentheses and (b) that $v_c + \bar{v}_T \approx 2v_c$

$$\frac{d\bar{v}_T}{dR} = -v_a\left(\frac{1}{R} - \frac{2}{h_\sigma}\right) + \frac{dv_c}{dR}.\quad (7)$$

Therefore, we find that

$$\Delta A \approx -\frac{v_a}{h_\sigma},\quad (8)$$

$$\Delta B \approx \frac{v_a}{R}\left(1 - \frac{R}{h_\sigma}\right).\quad (9)$$

These simple relations are to the author's knowledge not present in the literature. Olling & Dehnen (2003) use the Jeans equation to estimate $\Delta A$ and $\Delta B$, but they stop short of deriving the relation above[6]. Their value of $\Delta A = -0.14\,v_a$, where $v_a$ is in units of km s$^{-1}$ and $A$ in units of km s$^{-1}$ kpc$^{-1}$, computed for $h_\sigma = 0.9R_0$ agrees with equation (8), as in our units their value is $\approx 0.14 \times R_0 = 1.12$ and equation (8) gives $1/h_\sigma = 1.11$. Kuijken & Tremaine (1991) derived expressions for $\Delta A$ and $\Delta B$ to leading order in $\sigma_R^2/v_c^2$; it is straightforward to demonstrate that their expressions are almost equivalent to equations (8) and (9) (up to small terms).

Equation (8) explains why $A$ is robust to all changes to the DF except for $h_\sigma$: in the simple approximation $h_\sigma$ is the only parameter that affects $\Delta A$. From equation (8) we expect a linear relation between $\Delta A$ and $v_a$, that in the particular case of our fiducial model is one-to-one. We also expect $\Delta A$ to be negative in all cases. All of these analytical predictions are borne out in Figure 24.

Equation 9 shows why $\Delta B$ is much smaller than $\Delta A$: for our fiducial model, $\Delta B$ vanishes in this approximation. When $h_\sigma \neq R_0$, $\Delta B$ is much larger, as is seen in Figure 24. As for the fiducial model $\Delta B$ vanishes in the simple approximation, the actual (small) value of $\Delta B$ is determined by other factors such as the slope of the rotation curve (which affects $\gamma^2$) or the scale length. The sign of $\Delta B$ can therefore be both positive and negative. In the simple approximation above, $|\Delta B| < |\Delta A|$ as long as $h_\sigma < 2R_0$. While the velocity-dispersion profile in the Milky Way is likely quite flat (e.g., Bovy et al. 2012), $h_\sigma$ probably does not actually exceed $2\,R_0$, such that we should expect $\Delta B$ to be smaller than $\Delta A$. This implies that measurements of $B$ in the solar neighborhood are much less affected by the effects due to non-circular motions than determinations of $A$ (see also Olling & Dehnen 2003 who reach a similar conclusion).

---

[6] They do note that the $\Delta A$ and $\Delta B$ that they calculate are largely independent of $h_R$.

```
1  from galpy.df import quasiisothermaldf
2  from galpy.potential import MWPotential2014
3  from galpy.actionAngle import actionAngleStaeckel
4  # Setup actionAngle instance for action calcs
5  aAS= actionAngleStaeckel(pot=MWPotential2014,delta=0.45,
6                            c=True)
7  # Quasi-iso df w/ hr=1/3, hsr/z=1, sr(1)=0.2, sz(1)=0.1
8  df= quasiisothermaldf(1./3.,0.2,0.1,1.,1.,aA=aAS,
9                         pot=MWPotential2014)
10 # Evaluate DF w/ R,vR,vT,z,vz
11 df(0.9,0.1,0.8,0.05,0.02)
12 array([ 123.57158928])
13 # Evaluate DF w/ Orbit instance, return ln
14 from galpy.orbit import Orbit
15 df(Orbit([0.9,0.1,0.8,0.05,0.02]),log=True)
16 array([ 4.81682066])
17 # Evaluate DF marginalized over vz
18 df.pvRvT(0.1,0.9,0.9,0.05)
19 23.273310451852243
20 # Evaluate DF marginalized over vR,vT
21 df.pvz(0.02,0.9,0.05)
22 50.949586235238172
23 # Calculate the density
24 df.density(0.9,0.05)
25 12.73725936526167
26 # Estimate the DF's actual density scale length at z=0
27 df.estimate_hr(0.9,0.)
28 0.322420336223
29 # Estimate the DF's actual surface-density scale length
30 df.estimate_hr(0.9,None)
31 0.3805909132766462
32 # Estimate the DF's density scale height
33 df.estimate_hz(0.9,0.02)
34 0.064836202345657207
35 # Calculate the mean velocities
36 df.meanvR(0.9,0.05), df.meanvT(0.9,0.05),
37 df.meanvz(0.9,0.05)
38 (3.8432265354618213e-18, 0.90840425173325279,
39  -4.3579787517991084e-19)
40 # Calculate the velocity dispersions
41 from numpy import sqrt
42 sqrt(df.sigmaR2(0.9,0.05)), sqrt(df.sigmaz2(0.9,0.05))
43 (0.22695537077102387, 0.094215523962105044)
44 # Calculate the tilt of the velocity ellipsoid
45 df.tilt(0.9,0.05)
46 2.5166061974413765
47 # Calculate a higher-order moment of the velocity DF
48 df.vmomentdensity(0.9,0.05,6.,2.,2.,gl=True)
49 0.0001591190092366438
50 # Sample velocities at given R,z, check mean
51 vs= df.sampleV(0.9,0.05,n=500); mvt= numpy.mean(vs[:,1])
52 assert numpy.fabs(numpy.mean(vs[:,0])) < 0.05 # vR
53 assert numpy.fabs(mvt-df.meanvT(0.9,0.05)) < 0.01 #vT
54 assert numpy.fabs(numpy.mean(vs[:,2])) < 0.05 # vz
```

FIG. 25.— Methods of `quasiisothermaldf` instances, illustrated using a MWPotential2014 potential, with $\Sigma(R) \propto \exp[-R/(R_0/3)]$, $\sigma_R(R) = 0.2 \exp[-(R - R_0)/R_0]$, and $\sigma_z(R) = 0.1 \exp[-(R - R_0)/R_0]$. Other marginalizations similar to `df.pvRvT` and `df.pvz` are available as well.

### 6.3. *Three-dimensional distribution functions*

A three-dimensional disk DF that is included in `galpy` is the quasi-isothermal DF (qDF) of Binney (2010), including the improvements made by Binney & McMillan (2011). This qDF is a steady-state, axisymmetric DF that is a function of all three of the orbital actions $(J_r, J_z, J_\phi)$, where we denote the azimuthal action $J_\phi$ by $L_z$, as this action is equal to the $z$-component of the angular momentum for an axisymmetric potential (with rotational symmetry around the $\hat{\mathbf{z}}$ unit vector).

The quasi-isothermal DF is given by

$$\mathrm{qDF}(J_r, L_z, J_z) = f_{\sigma_R}(J_r, L_z) \times \frac{\nu}{2\pi\sigma_z^2} \, \exp\left(-\frac{\nu J_z}{\sigma_z^2(R_L)}\right), \quad (10)$$

where $f_{\sigma_R}$ is given by

$$f_{\sigma_R}(J_r, L_z) = \frac{\Omega n(R_L)}{\pi \sigma_R(R_L)^2 \kappa}\bigg|_{R_L} \times [1 + \tanh(L_z/L_0)]$$
$$\times \exp\left(-\frac{\kappa J_r}{\sigma_R^2(R_L)}\right). \quad (11)$$

The functions $\kappa$, $\Omega$, and $\nu$ are the epicycle, circular, and vertical frequencies (Binney & Tremaine 2008) evaluated at $R_L$, where $R_L$ is again the radius of the circular orbit with angular momentum $L_z$. We include the factor in equation (11) containing the tanh to eliminate stars on counter-rotating orbits following Binney & McMillan (2011), but there is also an initialization option to explicitly set the qDF to zero for counter-rotating orbits. The functions $n$, $\sigma_R$, and $\sigma_z$ are free functions of $R_L$, which indirectly determine the radial profiles of the tracer density, the radial velocity dispersion, and the vertical dispersion. However, these are merely *scale* profiles, as opposed to the actual, physical profiles that can be calculated by taking the appropriate moments of the DF (similar to the difference between the scale profiles and the actual profiles for the two-dimensional DFs above). In principle these three functions can take any form, but currently `galpy` only supports setting each of these to an exponential

$$n(R_L) \propto \exp(-R_L/h_R), \quad (12)$$
$$\sigma_R(R_L) = \sigma_{R,0} \exp(-[R_L - R_0]/h_{\sigma_R}), \quad (13)$$
$$\sigma_z(R_L) = \sigma_{z,0} \exp(-[R_L - R_0]/h_{\sigma_z}). \quad (14)$$

The qDF is available as `galpy.df.quasiisothermaldf`.

The `quasiisothermaldf` can be used with *any* Potential instance or list of such instances in `galpy`. The necessary action–angle calculations are performed using an instance of one of the subclasses of `actionAngle` that is provided at the initialization of the `quasiisothermaldf` instance. Any `actionAngle` instance can be used, but for the disk orbits that are represented by a qDF, primarily the `actionAngleAdiabatic` and `actionAngleStaeckel` as well as their gridded versions are useful.

The `quasiisothermaldf` implementation was used extensively in the analysis of Bovy et al. (2013). That paper contains some details of the implementation of some of the instance methods. In Figure 25 I provide examples of some of the instance methods to illustrate the use of `quasiisothermaldf` instances. These include methods to evaluate the qDF itself, the DF marginalized over one or two of the velocity components, and moments of the DF such as the density, mean velocities, and velocity dispersions. A method for sampling velocities from the DF at a given position is also included to aid in generating mock data from the qDF.

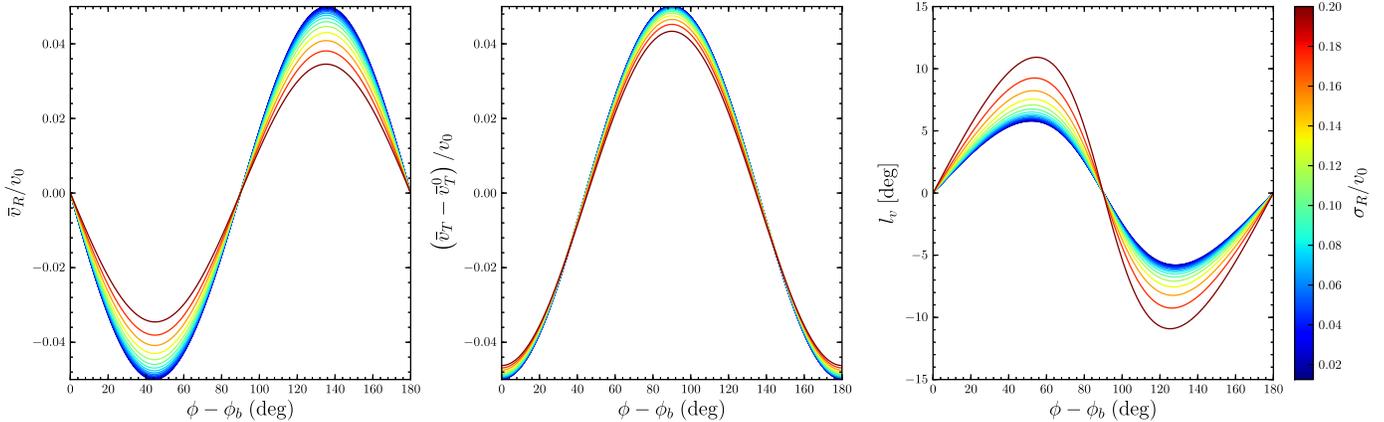

FIG. 26.— Stellar response of a kinematically-warm disk as a function of azimuth to an elliptical disk with an amplitude of 0.05 at $R_0$. An elliptical-disk perturbation $\phi(R, \phi) = 0.05/2 \cos 2(\phi - \phi_b)$ to the potential is slowly grown in an initially steady-state axisymmetric disk with a flat rotation curve and a Dehnen distribution function with $h_R = R_0/3$, $h_\sigma = R_0$, and radial velocity dispersion $\sigma_R$. The difference between the mean radial velocity $\bar{v}_R$, the mean rotational velocity $\bar{v}_T$, and the vertex deviation $l_v$ with respect to the axisymmetric value is shown (which is only non-zero for the mean rotational velocity). Different curves demonstrate the dependence of the response on $\sigma_R$. While the radial and rotational velocities respond less strongly for $\sigma_R > 0$ than for a cold disk, the vertex deviation is larger and becomes non-sinusoidal.

## 7. THE RESPONSE OF A KINEMATICALLY-WARM DISK TO NON-AXISYMMETRIC PERTURBATIONS

### 7.1. *Methodology*

To demonstrate the use of the `galpy` modules with a larger example, I describe in this section the implementation in `galpy` of a general method for calculating the response of the stellar disk to non-axisymmetric perturbations. This method was proposed by Dehnen (2000) and consists of assuming that the DF was axisymmetric and time-independent with a known DF $f_0$ some time in the past and that this initially-axisymmetric DF was then acted on by the non-axisymmetric perturbations. With these assumptions, the DF at a phase–space point $(R, \phi, v_R, v_T)$—all of the discussion here is limited to two-dimensional DFs in the mid-plane of a galaxy—can be evaluated by integrating an orbit launched at this phase–space point backward in time until the time at which the DF was axisymmetric was reached; at this time, the orbit is at the point $(R', \phi', v'_R, v'_T)$. By Liouville's theorem, we can then evaluate the DF $f_{\rm na}(R, \phi, v_R, v_T)$ today as

$$f_{\rm na}(R, \phi, v_R, v_T) = f_0(R', \phi', v'_R, v'_T). \qquad (15)$$

This simple procedure is implemented in the `evolveddiskdf` class in `galpy.df`. In its current implementation, it uses one of the `galpy.df.diskdfs`, i.e., `shudf` or `dehnendf` as $f_0$ and is therefore limited to power-law models of the axisymmetric part of the potential (see discussion in § 6.1). The expression in equation (15) allows for straightforward evaluation of the DF and its moments at a particular location, such as the mean velocities, velocity dispersions, and the vertex deviation. `evolveddiskdf` also contains methods to evaluate the Oort functions at a given position. In general, the Oort functions can be defined in terms of radial and azimuthal derivatives of the mean velocity (e.g., Kuijken & Tremaine 1994) and we therefore need to evaluate derivatives of $f_{\rm na}$. These are calculated directly us-ing the chain rule. For example,

$$\frac{\partial f_{\rm na}(R, \phi, v_R, v_T)}{\partial R} = \frac{\partial f_0(R', \phi', v'_R, v'_T)}{\partial R'} \frac{\partial R'}{\partial R}$$
$$+ \frac{\partial f_0(R', \phi', v'_R, v'_T)}{\partial v'_R} \frac{\partial v'_R}{\partial R} + \frac{\partial f_0(R', \phi', v'_R, v'_T)}{\partial v'_T} \frac{\partial v'_T}{\partial R}, \qquad (16)$$

where there is no $\partial f_0/\partial \phi'$ term, because the initial DF is axisymmetric. The expression for $\partial f_{\rm na}/\partial \phi$ is similar, replacing $\partial R$ with $\partial \phi$. The partial derivatives of $f_0$ with respect to $R, v_R$, and $v_T$ can be computed directly from the form of the initial DF. The derivatives $\partial R'/\partial R$, etc. are computed by integrating a small phase-space volume $(d\mathbf{x}, d\mathbf{v})$ along the orbit, using the `integrate_dxdv` Orbit method, described in § 4.2.

The methods of `evolveddiskdf` objects are similar to those of `diskdf` objects. One major difference is that $f_{\rm na}$ and its derivatives can be easily evaluated on a grid in $(v_R, v_T)$ at a given $(R, \phi)$, allowing for quick evaluation of the moments and Oort functions.

### 7.2. *Response to an $m = 2$ mode*

As an example of the functionality of `evolveddiskdf` objects, we investigate the response of a stellar disk to an non-rotating $m = 2$ mode (an elliptical perturbation; Kuijken & Tremaine 1994). This is a perturbation of the form

$$\phi(R, \phi) = \phi_0(R) + \frac{\epsilon(R) v_c^2(R)}{2} \cos 2(\phi - \phi_b), \qquad (17)$$

where $\phi_0(R)$ is the axisymmetric part of the potential, here modeled as a power-law or a logarithm (for a flat rotation curve). With this definition, the equipotential contours are approximately elliptical with axis ratio $1 - \epsilon$. When assuming a radial profile for the perturbation, we use $\epsilon(R) \propto R^{p-2\beta}$, where $\beta$ is the power-law exponent of the rotation curve. The fiducial model below has $p = \beta = 0$.

The response of a cold disk, i.e., one with $\sigma_R = 0$, is straightforward to calculate using the method in § 6.2.2 of Binney & Tremaine (2008). This was first done by

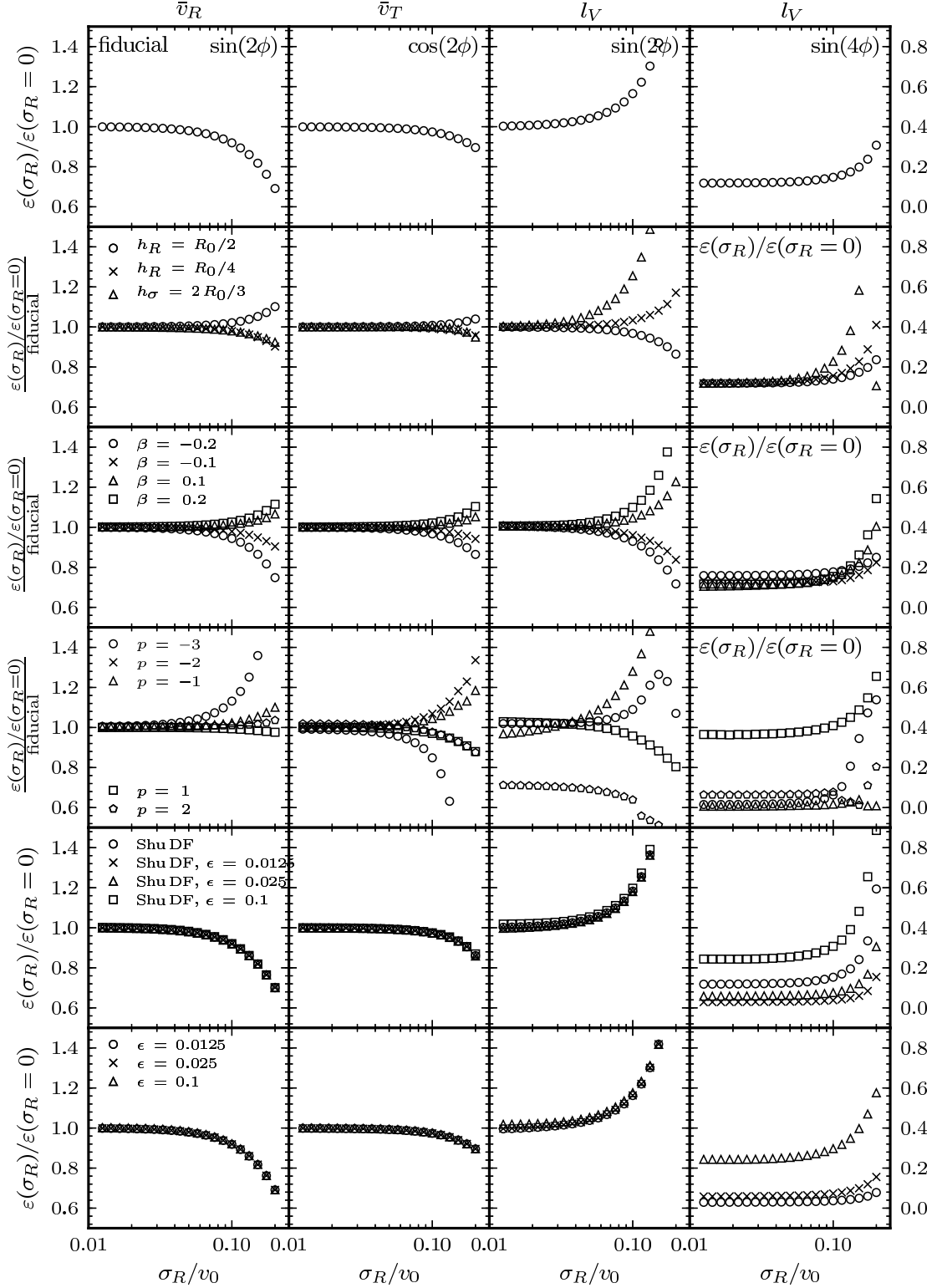

FIG. 27.— Detailed characterization of the response of a kinematically-warm stellar disk to an elliptical potential perturbation. This figure displays the main non-zero Fourier components of the mean velocity and vertex-deviation response as a function of velocity dispersion $\sigma_R$. The response is normalized by the response of a cold disk ($\sigma_R = 0$; see equations [18]–[20]). The mean radial-velocity and rotational-velocity response are well-described by a sine and cosine, respectively, of $2(\phi - \phi_b)$, while the vertex deviation has a substantial $\sin 4(\phi - \phi_b)$ component as well (shown in the last column; this column uses the tick marks on the right $y$ axis). All other Fourier components are negligible (that is, they amount to less than one percent of the cold response). The top panel displays the response for the fiducial model of Figure 26: a perturbation with constant ellipticity with radius, a flat rotation curve, and an initial Dehnen DF with $h_R = R_0/3$ and $h_\sigma = R_0$. The lower rows demonstrate the effect of varying the parameters of this model: the second row varies the DF parameters ($h_R$ and $h_\sigma$); the third row shows the effect of changing the shape of the rotation curve; the fourth row varies the radial profile of the ellipticity (which is $\propto R^p$). Rows two through four are normalized by the curves in the top row (except for the last column). The fifth row varies the form of the DF by showing the response for a Shu DF instead of a Dehnen DF as the initial DF; this is done for three different values of the magnitude of the perturbation. The bottom panel finally varies the magnitude of the perturbation for the Dehnen DF fiducial model. These two bottom rows demonstrate that the response is very close to linear in $\epsilon$.

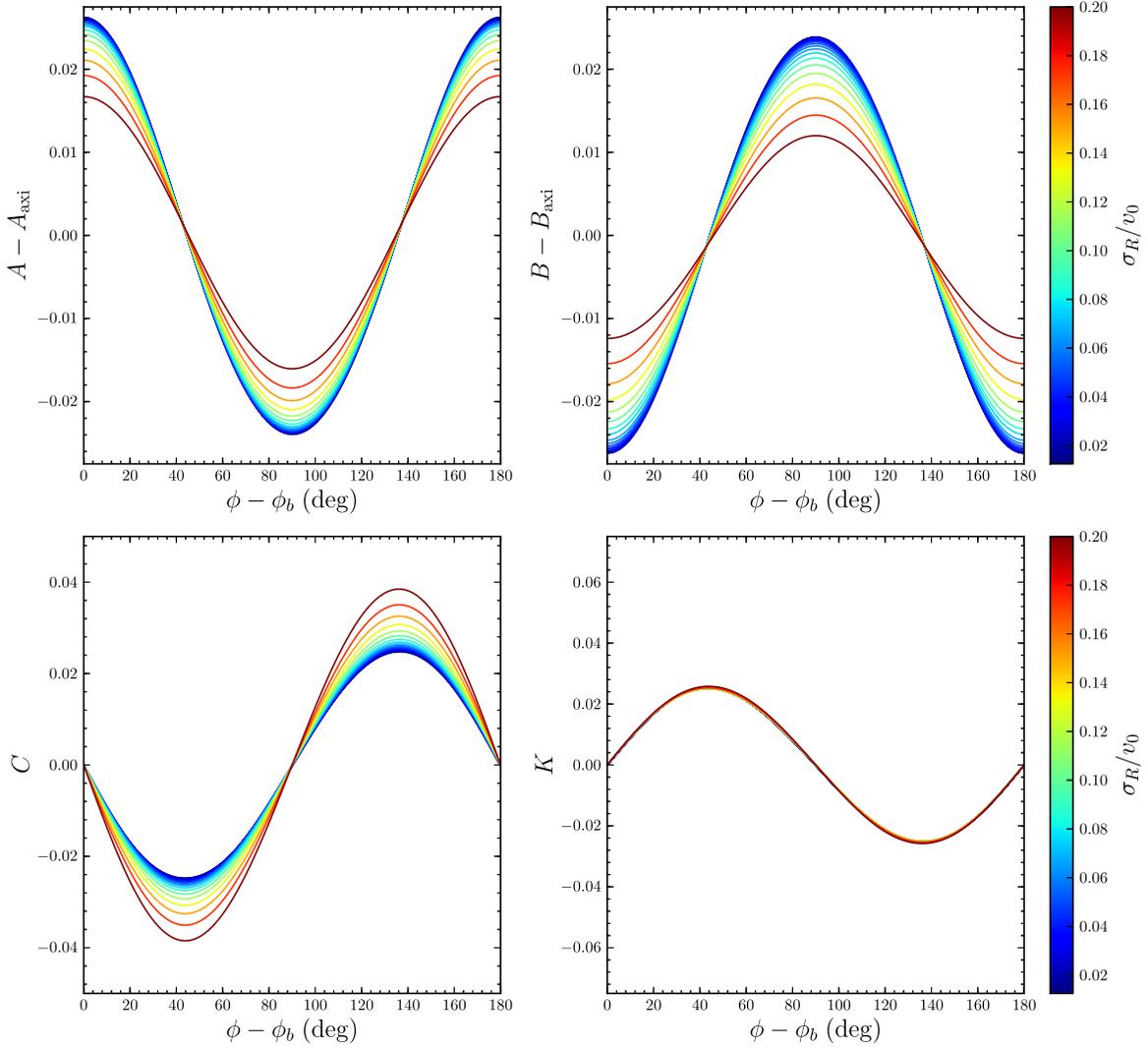

FIG. 28.— The effect of the same elliptical perturbation to the potential as in Figure 26 on the Oort functions measured at $R_0$, as a function of azimuth and initial velocity dispersion $\sigma_R$. The difference between the Oort functions and their axisymmetric value is shown ($C = K = 0$ for an axisymmetric disk). As for a cold disk, the Oort functions depend almost-perfectly sinusoidally on $\phi - \phi_b$, with $A$ and $B$ 90° out of phase and $C$ and $K$ −45° and 45° out of phase with respect to $A$. While $A$ and $B$ are closer to their axisymmetric values for warmer populations, $C$ responds more strongly the larger $\sigma_R$. Remarkably, $K$ is independent of $\sigma_R$.

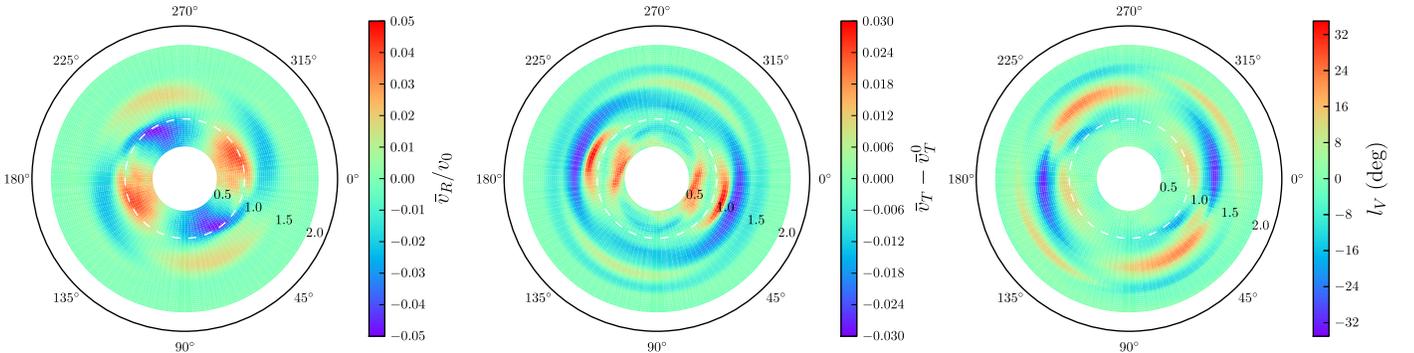

FIG. 29.— Response of a warm stellar disk to an adiabatically-grown quadrupole bar. The bar amplitude is grown over ≈ 41.5 bar periods and then held steady for $4\pi$ or two disk rotations at $R_0 = 1$. The background potential is logarithmic and the initial DF is a Dehnen DF with $h_R = R_0/3$, $h_\sigma = R_0$, and $\sigma_R = 0.2v_c$. The bar potential is of the form in equation (24) with $R_{\mathrm{OLR}} = 0.9R_0$ or $\Omega_b = 1.9\Omega_0$, a current angle between the major axis of the bar and the $\phi = 0$ direction of 25°, and a dimensionless bar amplitude of $\alpha = 0.01$. The three panels show the response in mean radial velocity $\bar{v}_R$, mean rotational velocity minus the axisymmetric mean ($\bar{v}_T - \bar{v}_T^0$), and the vertex deviation $l_v$. As expected, the response in all three has $\phi \to \phi + 180°$ symmetry. The mean velocity responds most strongly near the outer Lindblad resonance, indicated by the white dashed line, and at corotation, which lies at $R = 0.52R_0$. The bar can induce vertex deviations of typically 15° near the outer Lindblad resonance and around $R = 1.3R_0$.

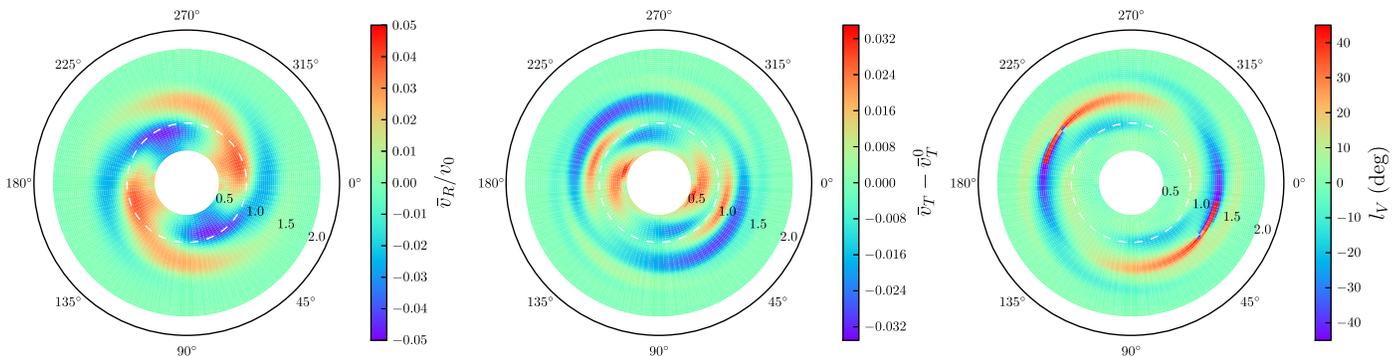

FIG. 30.— Same as Figure 29, but for a bar that is grown over two bar periods and evolved for two more bar periods (the model of Dehnen 2000). The rapid growth of the bar induces spiral-like features in the $\bar{v}_R$ response over a large part of the disk that are largely absent in the adiabatically-grown case. The amplitude of the response is approximately the same as in the adiabatic case.

Kuijken & Tremaine (1994) and we repeat the "cold response" here, as all of our $\sigma_R \neq 0$ results will be with respect to the cold response. The cold response is

$$\bar{v}_R = -\frac{1}{2}\left(\frac{p+2}{1-\beta}\right)\epsilon v_c \sin 2(\phi - \phi_b),\tag{18}$$

$$\bar{v}_T - v_c = -\left(\frac{1+p(1+\beta)/4}{1-\beta}\right)\epsilon v_c \cos 2(\phi - \phi_b),\tag{19}$$

$$l_v = \frac{(p+2)[4+(1+\beta)(3p-2\beta)]}{4(1-\beta)^2(1+2\beta)}\epsilon \sin 2(\phi - \phi_b),\tag{20}$$

where $l_v$ is the vertex deviation.

Figure 26 displays the result from a numerical calculation of the response at $R_0$ using the methods in `evolveddiskdf`. The fiducial model used in this figure is that of a flat perturbation ($p = 0$) with an amplitude of $\epsilon = 0.05$ in a background potential with a flat rotation curve ($\beta = 0$) that is slowly grown in an initially steady-state disk with a Dehnen DF with $h_R = R_0/3$ and $h_\sigma = R_0$. From equations (18)–(20), we can calculate that the cold response has an amplitude in $\bar{v}_R$, $\bar{v}_T - v_c$, and $l_v$ of $\epsilon$, $\epsilon$, and $2\epsilon = 5.73°$, respectively. The bluest line, which has $\sigma_R = 0.0125v_c$, is very close to this cold prediction, with the only exception that the vertex deviation does not behave exactly sinusoidally. The response of warmer populations is smaller than the cold response in $\bar{v}_R$ and $\bar{v}$ and larger in $l_v$. The response in $\bar{v}_T$ for populations with $\sigma_R > 0$ is measured against the mean rotational velocity $\bar{v}_T^0$ of the initial DF, i.e., taking into account the asymmetric drift. Figure 26 clearly demonstrates that the response in the mean velocities remains sinusoidal, while the $l_v$ response also keeps approximately the same shape (see below).

To further characterize the response of a warm disk to an elliptical perturbation, we have repeated the calculation displayed in Figure 26 for variations on the fiducial model. The results from this are shown in Figure 27. This figure shows the main Fourier modes of the response normalized by the cold response as a function of $\sigma_R$. As discussed above, for $\bar{v}_R$ and $\bar{v}_T$ the only significant modes are the $\sin 2(\phi - \phi_b)$ and $\cos 2(\phi - \phi_b)$ modes, respectively. For $l_v$ we find that we also need to include the $\sin 4(\phi - \phi_b)$ component, as this typically contributes about 10 %.

From linear perturbation theory, we expect the dependence of the response on $\sigma_R$ to be of the form $\mathcal{F} =$ $1 - f(\sigma_R/v_c)^2$ to lowest order in $\sigma_R$ and we fit this to the fiducial model. This gives

$$\mathcal{F}_{v_R} = 1 - 8(\sigma_R/v_c)^2,\tag{21}$$

$$\mathcal{F}_{v_T} = 1 - 2.5(\sigma_R/v_c)^2,\tag{22}$$

$$\mathcal{F}_{l_v} = 1 + 20(\sigma_R/v_c)^2,\tag{23}$$

where the index on $\mathcal{F}$ indicates the quantity that this factor applies to. For the mean velocities, reasonable variations in the DF and rotation curve only leads to corrections that are typically $\lesssim 10\%$; for $\bar{v}_R$ variations in the radial profile of the ellipticity parameterized by $p$ are also small. The warm response in $l_v$ is typically larger than the cold response and the magnitude of this increase depends more strongly on all of the parameters of the model. Its higher-order Fourier mode (that $\propto \sin 4[\phi - \phi_b]$) also strongly depends on the model and is not linear in $\epsilon$, as shown in the bottom two panels of the fourth column. Interpretations of the vertex deviation in terms of non-axisymmetric perturbations should therefore be treated with caution.

Finally, we also calculate the dependence of the Oort functions at $R_0$ on $\phi - \phi_b$ and $\sigma_R$. The cold response can be computed from equations (18)–(20) and is given explicitly in Kuijken & Tremaine (1994); we do not repeat it here. For the fiducial model, the response in all Oort functions has an amplitude of $\epsilon/2$ and the azimuthal dependence is as $\cos 2(\phi - \phi_b)$ for $A$ and $B$ and as $\sin 2(\phi - \phi_b)$ for $C$ and $K$. To calculate the warm response for $A$ and $B$ we again subtract the axisymmetric expectation, which is not exactly equal to the axisymmetric Oort constants that would be measured for a cold population (see § 6.2).

The result from this calculation is displayed in Figure 28. This figure demonstrates that the response for warmer populations stays sinusoidal, as expected from the sinusoidal velocity response of warm populations. The warm response in $A$ and $B$ is suppressed compared to the cold response and somewhat more strongly so for $B$. The warm response in $C$ is stronger than the cold response. Remarkably, the response in $K$ is independent of $\sigma_R$. Using the response factors in equations (21)–(23) and assuming that the only radial dependence of the warm-response/cold-response is in the $\sigma_R^2$ term, we can understand these trends. The warm-response factors for the Oort functions that one computes in this way are $\mathcal{F}_A = 1 - 8.5(\sigma_R/v_c)^2$, $\mathcal{F}_B = 1 - 18.5(\sigma_R/v_c)^2$,

$\mathcal{F}_C = 1 + 19(\sigma_R/v_c)^2$, and $\mathcal{F}_K = 1 - 13(\sigma_R/v_c)^2$. For $A$, $B$, and $C$ these agree reasonably well with the numerical calculations displayed in Figure 28, but this estimate entirely fails to explain the lack of $\sigma_R$ dependence for $K$. This is *not* due to unmodeled radial derivatives of $\mathcal{F}_R$, as from Figure 28 it can be seen that these would predict an even stronger suppression. It is likely due to higher-order terms in $\epsilon$, because these do not need a large prefactor to become important.

As discussed by Minchev et al. (2007) and Minchev & Quillen (2007), the dependence of the Oort constants on $\sigma_R$—foremost $C$ and $K$, as these are zero for an axisymmetric disk—can be used to constrain the type of non-axisymmetry perturbing the velocity field. Minchev & Quillen (2007) demonstrate that $|C|$ for spiral structure is smaller for warmer populations, while Minchev et al. (2007) find that $|C|$ is larger in the case of a bar (a rotating $m = 2$ perturbation). Here I have explicitly demonstrated that this is also the case for a non-rotating $m = 2$ mode. Tentative evidence that $|C|$ is larger for warmer populations was presented by Olling & Dehnen (2003). Gaia will allow the Oort functions to be measured with exquisite accuracy for various stellar populations and the tools in evolveddiskdf will be useful for determining the expected response and its dependence on $\sigma_R$ of different non-axisymmetric agents.

### 7.3. *Response to a weak bar*

As a final example of the use of evolveddiskdf, I investigate the response of the stellar disk to a weak bar. The bar model that we consider here is that of Dehnen (2000), who demonstrated that this model can explain the Hercules stream seen in the local velocity distribution. The response of the disk was further discussed in Mühlbauer & Dehnen (2003). I do not attempt as detailed a characterization as in the previous section on non-rotating $m = 2$ modes, but just show the response for a fiducial model.

The bar potential is modeled as a simple rotating quadrupole in DehnenBarPotential:

$$\Phi_b(R, \phi) = A_b(t) \cos[2(\phi - \Omega_b t)]$$
$$\times \begin{cases} -(R_b/R)^3, & \text{for } R \geq R_b, \\ (R/R_b)^3 - 2, & \text{for } R \leq R_b. \end{cases} \quad (24)$$

Here $\Omega_b$ is the pattern speed of the bar and $R_b$ is the bar radius, chosen to be 80 percent of the bar's corotation radius. The bar amplitude is parameterized by the dimensionless parameter $\alpha$, defined as

$$\alpha \equiv \frac{3 A_b}{v_c^2} \left( \frac{R_b}{R} \right)^3, \quad (25)$$

and this amplitude can be increased smoothly from zero to a given bar strength. We again use a Dehnen DF with parameters $h_R = R_0/3$ and $h_\sigma = R_0$ to model the initial state and a logarithmic background potential with $v_c(R) = v_0$. We only consider the case of a disk with $\sigma_R = 0.2\,v_c$. We consider the bar model with $\Omega_b = 1.9\Omega_0$, $\alpha = 0.01$, and a bar angle of $25°$.

Figure 29 shows the response in $\bar{v}_R$, $\bar{v}_T$, and $l_v$ over a large part of the disk when the bar is grown adiabatically. The radial velocity responds more strongly than the rotational velocity, with a maximum response of $0.04\,v_0$ for $\bar{v}_R$ versus a maximum of $0.03\,v_0$ for $\bar{v}_T$, both of which are obtained near the outer Lindblad resonance ($R_{\rm OLR} = 0.9\,R_0$). The vertex deviation is typically $\lesssim 15°$, with the largest values near the outer Lindblad resonance and near $R = 1.3\,R_0$.

Figure 30 shows the same as Figure 29, but for a bar that is grown over only two bar periods to mimic a more realistic bar growth scenario. The response is similar to the adiabatically-grown case, but the rapid bar growth induces spirality in the response, especially in $\bar{v}_R$. These figures can be compared to the results from Mühlbauer & Dehnen (2003), who used a forward-integration Monte-Carlo technique to calculate the moments of the velocity distribution. The agreement between these results and the results from Mühlbauer & Dehnen (2003) is good (note that the captions of their figures 6 and 7 appear to be switched).

The models shown here are useful for determining the influence of the bar on observations of the mean velocity field of stars in the Milky Way. Both the elliptical-disk and bar perturbations were used by Bovy et al. (2012) to estimate the effect of non-axisymmetry on their measurement of the Milky Way's rotation curve.

## 8. galpy DEVELOPMENT

I briefly discuss aspects related to the galpy codebase and its development in this section, to clarify how the code is structured, documented, and tested.

### 8.1. *Source code*

All of galpy's source code is public and has been so since it was first started, in July 2010. The source code is currently hosted in a git repository on GitHub at

http://github.com/jobovy/galpy .

There is no private development version. GitHub provides issue tracking that is used extensively to keep track of open issues and feature requests. Users are encouraged to report bugs through this issue tracker. As a git repository, it is easy for other users to copy the repository and create their own version of the code (and store it on GitHub); changes to the main repository are then merged through "pull requests", which is the preferred method for contributing to the codebase. galpy is released under the three-clause BSD license.

Overall, galpy consists currently of about 23,000 lines of code, with about 16,750 lines in python and 6,000 lines in C. The sizes of the major submodules are quite similar, with about 5,000 lines for each of the five main modules (actionAngle, df, orbit, potential, and util). While not incredibly large, a codebase of this size inevitably has hard-to-follow dependencies. Below I describe the automated testing framework that makes sure that changes to one part of the code do not break other parts.

galpy has an extensive set of documentation. This documentation is produced using Sphinx, which generates documentation in various format, and hosted by Read the Docs, which automatically builds the documentation upon each push to GitHub. Therefore, the documentation is always up to date. About 14,000 lines of documentation are contained in the source code, most of which are documentation of function inputs and outputs that are automatically placed in an API section of the online documentation by Sphinx. Apart from these,

`galpy` contains another $\approx 5,500$ lines of reStructured-Text in various tutorials. A PDF version of the online documentation currently weighs in at 283 pages.

## 8.2. *Automated testing and code coverage*

Much effort has been put into creating a useful and meaningful test suite to ensure the reliability of the `galpy` code and to maintain the integrity of the code as it is extended. To do this, `galpy` uses the `nose` test environment for `python`. The test suite, which we discuss in more detail below, is automatically run using the `Travis CI` continuous integration service. This online service detects pushes to the `GitHub` repository and downloads the code upon each push. `galpy`'s dependencies and `galpy` itself are then compiled and the test suite is run. This automatic procedure ensures that the code always compiles properly (at least on the `Travis CI` setup) and that all tests of the code are satisfied. The status of the code (whether it currently compiles and passes the test suite) can be checked on the galpy GitHub page.

The test suite currently consists of 500 test functions consisting of about 11,000 lines of code. These tests overall check the truth of about 20,000 assertions about the behavior of the code. All but 1,600 of these assertions are in tests of the potential and orbit-integration capabilities, because some of these tests are performed on grids of a large number of points.

Tests of `galpy.potential` functions are, for example, that the forces are correctly implemented as the negative derivatives of the potential and that the second derivatives of the potential are consistent with the forces (this is done using numerical differentiation). I also test that the Poisson equation is satisfied in cases where the density is implemented independently from the potential derivatives. All of these tests are automatically performed for all potentials contained in `galpy`. User-contributed new potentials that are registered in `galpy.potential` are also automatically subjected to these tests (without any need for the user to do anything). The potential test module further contains a variety of tests related to the mass, that the circular velocity and epicycle frequencies behave as expected, that the `MWPotential2014` of § 3.5 is implemented as specified in that section, etc.

Tests of the orbit-integration routines make up a significant chunk of the `galpy` test suite. I test energy conservation for all time-independent potentials and for all integrators and test that the Jacobi integral is conserved for the potentials with a fixed pattern speed. I further test that the energy error does not increase secularly for the symplectic integrators, by integrating orbits in a Kepler potential for 1,600 periods and asserting that the slope of a linear fit to the energy errors is close to zero. If I also test that Liouville's theorem is satisfied to test `integrate_dxdv`. Further there are various tests of the peri/apocenters, eccentricity, and maximum height of orbits in many of the potentials and of their analytical calculation, of the approximate conservation of the radial and vertical energy for orbits close to the galactic plane, of the interpolation of the orbit, and of the projection of the orbit in various coordinate and unit systems.

The methods of `actionAngle` instances are tested by checking that actions are conserved along orbits in static potentials, that the radial and vertical actions are very small for close-to-circular orbits, that the frequencies of close-to-circular orbits are approximately the epicycle frequencies (radial, rotational, and vertical), that the angles increase linearly in time, and that the slope of this increase is given by the frequencies. We further test that the action-angle coordinates computed by all methods agree for an isochrone potential, by checking that the actions, frequencies, and angles are the same when calculated by the method in question and `actionAngleIsochrone`. Thus, we know that all different methods are implemented consistently.

The methods in `galpy.df.diskdf` are tested by calculating moments of the DF and the Oort functions for very cold populations and comparing them to analytical predictions. This is done at a few different positions and for rotation curves that are flat, falling, and rising. We also check that the surface-density routines work as expected, both for cold and warm populations, and that the sampling methods return Monte Carlo samples that are consistent with the moments of the DF. The methods in `galpy.df.quasiisothermaldf` are also tested against analytical predictions for an axisymmetric DF and, for orbits close to the plane, against the calculations in `galpy.df.diskdf` (which should be similar). We also test that the sampling routine is consistent with the moments of the DF and that the methods that compute the DF marginalized over one or two velocity components are consistent with a simple Riemann sum of the DF. The methods in `galpy.df.evolveddiskdf` are primarily tested for potentials that are very close to axisymmetric: for such potentials the moments of the DF should agree with those of the initial `diskdf`. We also check the computations for an elliptical-disk perturbation applied to a very cold population against the analytical predictions (see § 7 above).

The utility functions in `galpy.util` are also well tested. All of the coordinate transformations contained in `bovy_coords` are tested for various points (such as the North Galactic Pole for `radec_to_lb`). The conversion factors in `bovy_conversion` are tested by checking that they scale correctly with the distance and velocity scales. The plotting routines in `bovy_plot` are not tested, because it is hard to meaningfully test plotting routines. Plotting routines in the rest of the `galpy` module are tested by making sure that they run without raising exceptions.

Finally, tests are run that all of the code examples given in this paper produce the output given here, such that these examples will remain valid in future updates to the code (or at least that any changes will be known).

The amount of the `galpy` code that is covered by all of these tests is tracked using coverage tools (`coverage.py` for the `python` code and `gcov` for the C code) and automatically reported online after `Travis CI` runs the test suite. The coverage of the code is currently 99.6 %, that is, 14,166 of 14,220 relevant lines are touched by the test code (relevant lines here indicates individual statements and combines multi-line statements into one relevant line; this is one of the the main reasons for the discrepancy between this number of lines and the number of 23,000 lines of code quoted above for the whole galpy module). All python files are covered at more than 99 %. The only lines that are not covered in the whole codebase are related to special cases that are very unlikely to come up in practice and paths through the code that are currently impossible to access through the user interface (but are left for completeness).

## 9. FUTURE OUTLOOK

`galpy` is under continuous development and will be extended with more potentials, action-angle methods, distribution functions, etc. in the future. One extension that has not been discussed here is a dynamical modeling framework for tidal streams, contained in `galpy.df.streamdf`, that is described in detail in Bovy (2014). Further support for dealing with tidal streams will be added soon. Other DFs that will likely be added in the near future are DFs for spheroidal systems, such as the local stellar halo, to aid in dynamical modeling of the Gaia data.

One major extension that has already started being developed is extensive support for dealing with the outputs from *N*-body simulations. Work has been started, primarily by Rok Roškar, on implementing a `SnapshotPotential` class that allows the potential of a snapshot from an *N*-body simulation to be evaluated at an arbitrary point using direct summation. Such an instance can be used by an `InterpSnapshotPotential`—a subclass of `interpRZPotential`—to tabulate the snapshot's potential on a grid. This instance of an interpolated potential can then be used anywhere in `galpy` like any other potential. For example, one can integrate test-particle orbits in this static snapshot potential, evaluate the integrals of the motion, or the actions.

Further small and large possible extensions of the `galpy` codebase are described on `galpy`'s GitHub page.

It is a pleasure to thank Mark Fardal and Denis Erkal for some direct contributions to the code, Eduardo Balbinot, Dylan Gregersen, Željko Ivezić, Resmi Lekshmi, Wai-Hin Ngan, Timothy Pickering, Rok Roškar, and Wilma Trick for bug reports, and the anonymous referee for helpful comments. J.B. was supported by NASA through Hubble Fellowship grant HST-HF-51285.01 from the Space Telescope Science Institute, which is operated by the Association of Universities for Research in Astronomy, Incorporated, under NASA contract NAS5-26555. This research was also partially supported by the National Science Foundation (grant AST-0908357) and by the German Research Foundation DFG through grant SFB 881 (A3). This work was further made possibly by the open-source codes numpy (Oliphant 2006), scipy (Jones et al. 2001), matplotlib (Hunter 2007), and the GNU Scientific Library (Galassi et al. 2009). Version-control, continuous-integration, test-statistics, and documentation services are provided by GitHub, Travis CI, Coveralls, and Read the Docs and are gratefully acknowledged.

## APPENDIX

### COORDINATE TRANSFORMATIONS IN GALPY

`galpy` contains a number of utility modules to help with sampling one-dimensional distribution functions (`galpy.util.bovy_ars`), plotting (`galpy.util.bovy_plot`), unit conversions (see § 2.2; `galpy.util.bovy_conversion`) and coordinate transformations (`galpy.util.bovy_coords`). We discuss the latter here briefly, because it is generally useful and contains some transformations not commonly found in coordinate-transformation `python` packages.

The coordinate transformations supported by `galpy` are those between (a) equatorial, (b) Galactic (both spherical and rectangular), and (c) Galactocentric coordinates (rectangular and cylindrical). Transformations for both positions and velocities in these different coordinate systems are implemented. All transformations work for scalar and array input (for multiple points) and only rely on `numpy` functions; array operations are therefore very fast.

Celestial positions and proper motions can be converted between equatorial and Galactic coordinates using the functions `radec_to_lb`, `pmrapmdec_to_pmllpmbb`, and similar functions for the inverse transformations. Spherical Galactic coordinates can be transformed to rectangular coordinates using `lbd_to_XYZ`, `vrpmllpmbb_to_vxvyvz`, `sphergal_to_rectgal`, and the inverse transformations are implemented under similar names. Angles are input in degrees or radians, distances in kpc, proper motions in $\mathrm{mas\,yr^{-1}}$, and line-of-sight velocities in $\mathrm{km\,s^{-1}}$. `galpy` can also convert uncertainties in the equatorial proper motions to uncertainties in Galactic proper motions using `cov_pmrapmdec_to_pmllpmbb` and can transform the uncertainty covariance matrix in Galactic coordinates to that of the velocities in the rectangular frame with `cov_dvrpmllbb_to_vxyz`.

As `galpy` primarily works in Galactocentric cylindrical coordinates, Galactic coordinates can be further transformed to the Galactocentric restframe. This functionality is contained in the functions `XYZ_to_galcencyl`, `XYZ_to_galcenrect`, `vxvyvz_to_galcenrect`, `vxvyvz_to_galcencyl`, and similar inverse transformations. To make use of these transformations, the user needs to specify the Sun's position and velocity with respect to the Galactic center. A typical workflow to transform an observed position and velocity to the Galactocentric cylindrical frame is

```python
from galpy.util import bovy_coords
ra, dec, dist= 161., 50., 8.5
pmra, pmdec, vlos= -6.8, -10., -115.
# Convert to Galactic and then to rect. Galactic
ll, bb= bovy_coords.radec_to_lb(ra,dec,degree=True)
pmll, pmbb= bovy_coords.pmrapmdec_to_pmllpmbb(pmra,pmdec,ra,dec,degree=True)
X,Y,Z= bovy_coords.lbd_to_XYZ(ll,bb,dist,degree=True)
vX,vY,vZ= bovy_coords.vrpmllpmbb_to_vxvyvz(vlos,pmll,pmbb,X,Y,Z,XYZ=True)
# Convert to cylindrical Galactocentric
```

```
# Assuming Sun's distance to GC is (8,0.025) in (R,z)
R,phi,z= bovy_coords.XYZ_to_galcencyl(X,Y,Z,Xsun=8.,Zsun=0.025)
vR,vT,vz= bovy_coords.vxvyvz_to_galcencyl(vX,vY,vZ,R,phi,Z,vsun=[-10.1,244.,6.7],galcen=True)
print R, phi, z
12.5132851516 0.121774090734 7.12412823549
print vR, vT, vz
78.961682923 -241.492477724 -102.839654422
```

As also discussed in § 8.2, the `bovy_coords` functions have 100% test coverage from 30 test functions and 272 test assertions (individual tests).

## REFERENCES

Binney, J. & Tremaine, S. 2008, Galactic Dynamics: Second Edition
Binney, J. J. 2010, MNRAS, 401, 2318
Binney, J. J. & McMillan, P. 2011, MNRAS, 413, 1889
Binney, J. J. 2012, MNRAS, 426, 1324
Bovy, J. 2014, ApJ, in press
Bovy, J., Allende Prieto, C., Beers, T. C., et al. 2012, ApJ, 759, 131
Bovy, J. & Rix, H.-W. 2013, ApJ, 779, 115
Bovy, J., & Tremaine, S. 2012, ApJ, 756, 89
Clemens, D. P. 1985, ApJ, 295, 422
Dehnen, W. & Binney, J. 1998, MNRAS, 294, 429
Dehnen, W. 1999, AJ, 118, 1201
Dehnen, W. 2000, AJ, 119, 800
Dexter, J. & O'Leary, R. 2013, ApJ, 783, L7
Dormand, J. R & Prince, P. J. 1980, J. Comp. Appl. Math., 6, 19
Forest, E. & Ruth, R. D. 1990, Physica D Nonlinear Phenomena, 43, 105
Galassi, M., et al. 2009, GNU Scientific Library Reference Manual (3rd Ed.), ISBN 0954612078
Gillessen, S., Eisenhauer, F., Trippe, S., et al. 2009, ApJ, 692, 1075
Holmberg, J. & Flynn, C. 2000, MNRAS, 313, 209
Hindmarsh, A. C., 1983, in Scientific Computing, R. S. Stepleman et al. (eds.), North-Holland, Amsterdam, 1, 55
Hunter, J. D. 2007, Computing In Science & Engineering, 9, 90
Jones, E., Oliphant, T., & Peterson, P. 2001, http://www.scipy.org/
Kinoshita, H., Yoshida, H., & Nakai, H. 1991, CeMDA, 50, 59
Kuijken K. & Tremaine, S. 1991, in Dynamics of Disk Galaxies, ed. B. Sundelius (Göteborg: Göteborg Univ. Press), 257
Kuijken K. & Tremaine, S. 1994, ApJ, 421, 178
McClure-Griffiths, N. M. & Dickey, J. M. 2007, ApJ, 671, 427
Minchev, I., Nordhaus, J., & Quillen, A. C. 2007, ApJ, 664, L31
Minchev, I. & Quillen, A. C. 2007, MNRAS, 377, 1163
Mühlbauer, G. & Dehnen, W. 2003, A&A, 401, 975
Navarro, J. F., Frenk, C. S., & White, S. D. M. 1997, ApJ, 490, 493
Oliphant, T. 2006, A Guide to NumPy, 1 (Trelgol Publishing USA)
Olling, R. P. & Dehnen, W. 2003, ApJ, 599, 275
Perryman, M. A. C., et al. 2001, A&A, 369, 339
Piffl, T., Binney, J., McMillan, P. J., et al. 2014, MNRAS, submitted
Sanders, J. 2012, MNRAS, 426, 128
Sharma, S. & Bland-Hawthorn, J. 2013, ApJ, 773, 183
Shu, F. H. 1969, ApJ, 158, 505
Smith, M., Ruchti, G. R., Helmi, A., et al. 2007, 379, 755
Springel, V. 2005, MNRAS, 364, 1105
Teuben, P. J. 1995, in Astronomical Data Analysis Software and Systems IV, ed. R. Shaw, H. E. Payne and J. J. E. Hayes., PASP Conf Series 77, 398
Xue, X. X., Rix, H.-W., Zhao, G., et al. 2008, ApJ, 684, 1143
Zhang, L., Rix, H.-W., van de Ven, G., Bovy, J., Liu, C., Zhao, G. 2013, ApJ, 772, 108